%% file: main.tex
\documentclass[
11pt, % The default document font size, options: 10pt, 11pt, 12pt
english, % ngerman for German
singlespacing, % Single line spacing, alternatives: onehalfspacing or doublespacing
headsepline, % Uncomment to get a line under the header
chapterinoneline, % Uncomment to place the chapter title next to the number on one line
]{MastersDoctoralThesis} % The class file specifying the document structure

\usepackage[utf8]{inputenc} % Required for inputting international characters
\usepackage[T1]{fontenc} % Output font encoding for international characters

\usepackage{mathpazo} % Use the Palatino font by default

\usepackage[style=ieee,natbib=true]{biblatex} % Use the bibtex backend with the authoryear citation style (which resembles APA)

\usepackage[autostyle=true]{csquotes} % Required to generate language-dependent quotes in the bibliography

\RequirePackage{amsmath}
\RequirePackage{amsthm}
\RequirePackage{dsfont}

\let\originalleft\left
\let\originalright\right
\renewcommand{\left}{\mathopen{}\mathclose\bgroup\originalleft}
\renewcommand{\right}{\aftergroup\egroup\originalright}

\DeclareMathOperator*{\argmax}{arg\,max}

\newcommand{\real}  {\ensuremath{\mathds{R}}}

\newcommand{\diff}[1]{\ensuremath{\operatorname{d}\!{#1}}}

\newcommand{\SNR}{\text{SNR}}

\RequirePackage{subcaption}

\usepackage{multirow}

\thesistitle{Probabilistic Shaping for the AWGN Channel} % Your thesis title, this is used in the title and abstract, print it elsewhere with \ttitle
\supervisor{Prof. Frank \textsc{Kschischang}} % Your supervisor's name, this is used in the title page, print it elsewhere with \supname
\examiner{} % Your examiner's name, this is not currently used anywhere in the template, print it elsewhere with \examname
\degree{} % Your degree name, this is used in the title page and abstract, print it elsewhere with \degreename
\author{Sébastien \textsc{Delsad}} % Your name, this is used in the title page and abstract, print it elsewhere with \authorname
\addresses{} % Your address, this is not currently used anywhere in the template, print it elsewhere with \addressname

\subject{Information Theory} % Your subject area, this is not currently used anywhere in the template, print it elsewhere with \subjectname
 % Keywords for your thesis, this is not currently used anywhere in the template, print it elsewhere with \keywordnames
\university{\href{https://www.epfl.ch/en/}{EPFL}} % Your university's name and URL, this is used in the title page and abstract, print it elsewhere with \univname
\department{\href{https://www.utoronto.ca/}{University of Toronto}} % Your department's name and URL, this is used in the title page and abstract, print it elsewhere with \deptname
\group{\href{https://www.epfl.ch/en/}{EPFL}} % Your research group's name and URL, this is used in the title page, print it elsewhere with \groupname
\faculty{} % Your faculty's name and URL, this is used in the title page and abstract, print it elsewhere with \facname

\AtBeginDocument{
\hypersetup{pdftitle=\ttitle} % Set the PDF's title to your title
\hypersetup{pdfauthor=\authorname} % Set the PDF's author to your name
\hypersetup{pdfkeywords=\keywordnames} % Set the PDF's keywords to your keywords
}

\begin{document}

\frontmatter % Use roman page numbering style (i, ii, iii, iv...) for the pre-content pages

\pagestyle{plain} % Default to the plain heading style until the thesis style is called for the body content

%----------------------------------------------------------------------------------------
%	TITLE PAGE
%----------------------------------------------------------------------------------------

\begin{titlepage}
\begin{center}

\vspace*{.06\textheight}
{\scshape\LARGE \univname\par}\vspace{0.5cm} % University name
{\scshape\LARGE \deptname\par}\vspace{1.5cm} % University name
\textsc{\Large Semester Project}\\[1cm] % Thesis type

\HRule \\[0.4cm] % Horizontal line
{\huge \bfseries \ttitle\par}\vspace{0.4cm} % Thesis title
\HRule \\[2cm] % Horizontal line
 
\begin{minipage}[t]{0.4\textwidth}
\begin{flushleft} \large
\emph{Author:}\\
\href{https://linkedin.com/in/sébastien-delsad-32b743239}{\authorname} % Author name - remove the \href bracket to remove the link
\end{flushleft}
\end{minipage}
\begin{minipage}[t]{0.4\textwidth}
\begin{flushright} \large
\emph{Supervisor:} \\
\href{https://www.comm.utoronto.ca/frank/}{\supname} % Supervisor name - remove the \href bracket to remove the link  
\end{flushright}
\end{minipage}\\[3cm]
 
\vfill

{\large December 2022}\\[4cm] % Date

\vfill
\end{center}
\end{titlepage}

%----------------------------------------------------------------------------------------
%	ABSTRACT PAGE
%----------------------------------------------------------------------------------------

\begin{abstract}
\addchaptertocentry{\abstractname} % Add the abstract to the table of contents
In this report, we study communication over an additive white Gaussian noise channel with a fixed signal constellation. We measure how much information we can send through this channel and how to improve the rate of communication by changing the input probability distribution. More precisely, we study the mutual information obtained from the Maxwell--Boltzmann distribution, the Blahut--Arimoto algorithm and a constrained version of the Blahut--Arimoto algorithm. We emphasise the fact that the Maxwell--Boltzmann distribution is not optimal. We also observe that the Blahut--Arimoto algorithm does not give us the best mutual information over SNR. To get the optimal distribution for a fixed SNR, we have to implement a constrained version of the Blahut--Arimoto algorithm.
\end{abstract}

%----------------------------------------------------------------------------------------
%	LIST OF CONTENTS/FIGURES/TABLES PAGES
%----------------------------------------------------------------------------------------

\tableofcontents % Prints the main table of contents

%\listoffigures % Prints the list of figures

%\listoftables % Prints the list of tables

%----------------------------------------------------------------------------------------
%	THESIS CONTENT - CHAPTERS
%----------------------------------------------------------------------------------------

\mainmatter % Begin numeric (1,2,3...) page numbering

\pagestyle{thesis} % Return the page headers back to the "thesis" style

% Include the chapters of the thesis as separate files from the Chapters folder
% Uncomment the lines as you write the chapters

\include{Chapters/1Introduction}
\include{Chapters/2MI_Continuous}
\include{Chapters/3MI_Discrete}
\include{Chapters/4MB_Distribution} 
\include{Chapters/5BA_Distribution}
\include{Chapters/6BA_Constrained}
\include{Chapters/7Conclusion.tex}

%----------------------------------------------------------------------------------------
%	THESIS CONTENT - APPENDICES
%----------------------------------------------------------------------------------------

\appendix % Cue to tell LaTeX that the following "chapters" are Appendices

% Include the appendices of the thesis as separate files from the Appendices folder
% Uncomment the lines as you write the Appendices

%\include{Appendices/AppendixA}
%\include{Appendices/AppendixB}
%\include{Appendices/AppendixC}

%----------------------------------------------------------------------------------------
%	ABBREVIATIONS
%----------------------------------------------------------------------------------------

\begin{abbreviations}{ll} % Include a list of abbreviations (a table of two columns)

\textbf{AWGN} & \textbf{A}dditive \textbf{W}hite \textbf{G}aussian \textbf{N}oise\\
\textbf{SNR} & \textbf{S}ignal-to-\textbf{N}oise \textbf{R}atio\\
\textbf{RV} & \textbf{R}andom \textbf{V}ariable\\
\textbf{MI} & \textbf{M}utual \textbf{I}nformation\\
\textbf{QAM} & \textbf{Q}uadrature \textbf{A}mplitude \textbf{M}odulation\\
\textbf{PSK} & \textbf{P}hase \textbf{S}hift \textbf{K}eying\\
\textbf{AMPM} & \textbf{A}mplitude \textbf{M}odulation and \textbf{P}hase \textbf{M}odulation \\
\textbf{PAM} & \textbf{P}ulse \textbf{A}mplitude \textbf{M}odulation\\
\textbf{MB} & \textbf{M}axwell--\textbf{B}oltzmann\\
\textbf{BA} & \textbf{B}lahut--\textbf{A}rimoto\\
\textbf{KL} & \textbf{K}ullback--\textbf{L}eibler\\

\end{abbreviations}

%----------------------------------------------------------------------------------------
%	ACKNOWLEDGEMENTS
%----------------------------------------------------------------------------------------

\begin{acknowledgements}
\addchaptertocentry{\acknowledgementname} % Add the acknowledgements to the table of contents
I am very grateful to Dr. Reza Rafie for introducing me to the concept of shaping
and for his precious help throughout the semester. I would like to extend my sincere thanks to
Prof. Frank Kschischang for his invaluable patience and feedback. I would also like to thank my parents for their support throughout my studies.
\end{acknowledgements}

%----------------------------------------------------------------------------------------
%	BIBLIOGRAPHY
%----------------------------------------------------------------------------------------

\printbibliography[heading=bibintoc]

%----------------------------------------------------------------------------------------

\end{document}

%% file: Chapters/1Introduction.tex
\chapter{Introduction} % Main chapter title

\begin{figure}[h]
    \centering
    \includegraphics[width=10cm]{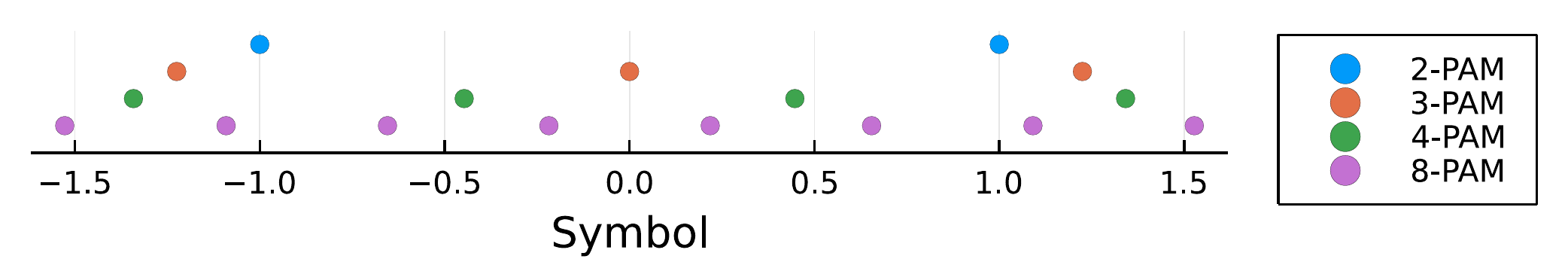}
    \caption{Examples of PAM constellations}
    \label{fig:PAMConstellationExamples}
\end{figure}

\begin{figure}[ht]
    \centering
    \includegraphics[width=10cm]{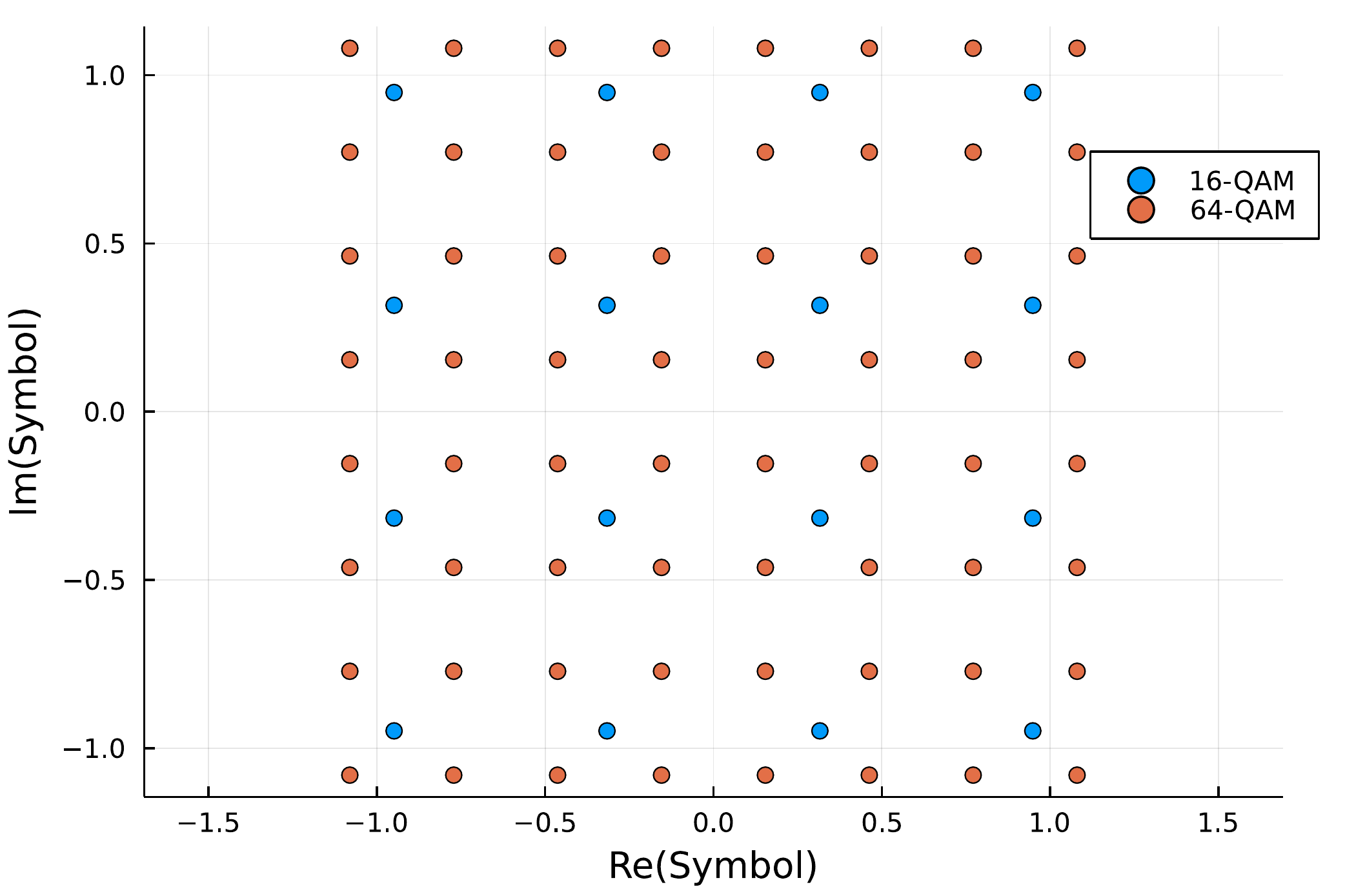}
    \caption{Examples of QAM constellations}
    \label{fig:QAMConstellationExamples}
\end{figure}

When a signal is transmitted over a noisy channel, it will experience some alteration. It will be modified by the channel it traverses. We consider a discrete-time Additive White Gaussian Noise (AWGN) channel. It is well-known that this channel can represent a bandwidth-limited continuous-time AWGN channel \cite{gallager2008principles}. For us, the AWGN channel is represented by a series of outputs $Y_i$ at discrete time event index $i$. The output $Y_i$ is the sum of the input $X_i$ and noise $W_i$, where $W_i$ is independent, identically distributed and drawn from a zero-mean normal distribution with standard deviation $\sigma$. The equation that relates the input and output for the AWGN channel is
\begin{equation*}
    Y_i = X_i + W_i .
\end{equation*}
Two cases are considered: complex-input and real-input AWGN channels. The distinction should be clear from the context.
\\\\
It would be interesting to have a measure of how “bad” a channel is. In other words, we would like to know how much noise the channel adds compared to the energy of our input signal. To this end, we use Signal power over Noise power Ratio (SNR). First, we have to define what is the power of a signal.
\\
Let $S$ be a random variable. We define its power to be the expected value of the squared magnitude of the outcomes of the random variable:
\[
\mathcal{P} = \mathbb{E}[|S|^2] .
\]
We now define the SNR as the ratio of the signal power to the noise power:
\[
\SNR = \frac{\mathcal{P}_{signal}}{\mathcal{P}_{noise}} = \frac{\mathbb{E}[|X_i|^2]}{\mathbb{E}[|W_i|^2]} .
\]
We can see that a ratio higher than 1:1 means that the power of the signal is higher than the power of the noise. Intuitively, the higher the SNR is, the better the channel behaves.
Note that we often express the SNR in decibel scale that is $\SNR_{dB} = 10\log_{10}(\SNR)$.
\\
\\
Now that we can measure how much a channel modifies our input, we would want to measure how much information we can pass on average over the channel.
To this end, we calculate the capacity of the channel. But first, let's define the concept of entropy and the mutual information between two random variables. They will later help us define the capacity of a channel.
\\
Conceptually, the entropy is a measure of disorder and uncertainty. In our case, we will use most of the time the binary entropy, a special case of the entropy, which measures the quantity of information in bits. We assume $X$ takes values in the alphabet $\Omega$ and is distributed according to $p: \Omega \to [0,1]$. We define the entropy as follows:
\[
\mathrm{H}_k(X) = \mathbb{E}[-\log_k(p(X))] .
\]
Note that it does not depend on the actual values of the events $x_i \in \Omega$, but only on the probability that $x_i$ occurs.
In the case of binary entropy, $k$ equals two which we simply denote as $H(X)$.
\\\\
The mutual information between two random variables X and Y tells us how much information we can have about X when knowing Y. It measures the statistical dependence between the two random variables. The mutual information is defined as
\begin{equation}
I(X, Y) = \mathbb{E}_{X,Y}\left[\log_2\left(\frac{p(X, Y)}{p(X)p(Y)}\right)\right],
\label{eqn:mutualInformation}
\end{equation}
where $p(x, y)$ expresses the joint probability that the events $x$ and $y$ occur together.
Note that we can also express the mutual information as the difference between the entropy of $Y$ and the entropy of $Y$ knowing $X$: 
\begin{equation}
    I(X,Y) = H(Y) - H(Y|X) = H(X) - H(X|Y) ,
    \label{eqn:MI as difference of entropy}
\end{equation}
which makes easier to interpret what the mutual information is. As we are using the base two logarithm and the binary entropy, the mutual information is expressed in bits.
\\
We can now define the capacity of a fixed input alphabet. Note that we will sometimes call the input alphabet a constellation.
Let $\Omega$ be an input alphabet contained in a vector space of $m$ dimensions. Let $X$ be a random variable such that $X$ takes values in $\Omega$ and let Y be the random variable representing the output of the memoryless channel. The capacity is then
\[
C = \sup\limits_{p_X(x)}I(X, Y) ,
\]
where $p_X(x)$ denotes the probability that $X$ takes value $x \in \Omega$. We can see $p_X(x)$ as the probability of sending the symbol $x$. The capacity of a channel is the highest possible mutual information between $X$ and $Y$ for a given fixed constellation.\\
The Shannon--Hartley theorem gives an exact value for the capacity of the AWGN channel knowing its SNR. The theorem tells us that 
\[ 
C = \log_2(1 + \SNR) \text{  [bits / 2 dimensions]} .
\]
Reliable communication is impossible if the rate of communication is bigger than this fundamental bound. \\
It is important to note the difference between the capacity of a constellation and the capacity of the AWGN channel. The later gives us the maximum mutual information achievable over all constellations and over all input distribution. The capacity of a constellation gives us the maximum mutual information achievable for a given fixed constellation.
\\\\
In this report, we will study different probability distribution and in order to compare them, we will use the Kullback--Leibler divergence. It is a measure of similarity between two probability distributions. Let $P$ and $Q$ be two probability distributions defined on the same probability space $\mathcal{X}$. The KL divergence between $P$ and $Q$ is defined to be:
\[
    D_{KL}(P || Q) = \sum \limits_{x \in \mathcal{X}} P(x)\log(\frac{P(x)}{Q(x)}) .
\]
Note that the KL divergence is not commutative. We define a measure similar to the KL distance that is commutative as:
\[
    D^c_{KL}(P, Q) = \frac{D_{KL}(P || Q) + D_{KL}(Q || P)}{2} .
\]
If the two probability distributions are very similar, their KL divergence is close to zero.
\\\\
Throughout this report, we will use some well known constellations. A constellation is an input alphabet. In one dimension, we will use the PAM constellations. PAM stands for Pulse Amplitude Modulation. Figure \ref{fig:PAMConstellationExamples} gives examples of PAM constellations that have been scaled in order to have unit energy.
\\
Most of the time, however, we will use 2-D constellations. Some well known types of constellations are the PSK constellations which stand for Phase Shift Keying and the QAM constellations which stand for Quadrature Amplitude Modulation. We will also use the AMPM constellations which stand for Amplitude Modulation and Phase Modulation. In Figure \ref{fig:QAMConstellationExamples}, the constellations of unit energy 16-QAM and 64-QAM are shown.
\\
Our goal in this report will be to first understand the concepts introduced by trying to reproduce the results of the Section II of \cite{channelCoding}. Then, we will try to modify the probability mass function of the input alphabet in order to maximise the mutual information. This process is called probabilistic shaping.

%% file: Chapters/2MI_Continuous.tex
%%%%%%%%%%%%%%%%%%%%%%%%%%%%%%%%%%%%%%%%%%%%%%%%%%%%%%%%%%%%%%%%%%%%%%%%
\chapter{Mutual Information for a Continuous-Output Channel}
%%%%%%%%%%%%%%%%%%%%%%%%%%%%%%%%%%%%%%%%%%%%%%%%%%%%%%%%%%%%%%%%%%%%%%%%

In this section, we will estimate the mutual information of the AWGN channel for a fixed finite constellation and a fixed input distribution. The goal is to reproduce the results of Section II of \cite{channelCoding} and to generalise to inputs having a specific distribution.

\section{MI of an AWGN Channel with Uniform-Input Distribution}

We consider a discrete random variable $X$ which takes values in the finite alphabet $\Omega \subseteq V$, where V is a vector space of dimension $m$. Let $n$ be the number of symbols in the constellation, $n = |\Omega|$. The random variable $X$ is distributed according to $p_X: X \to [0,1]$. Let $Y$ be a continuous random variable which models the output of the AWGN channel. The random variable $Y$ is distributed according to probability density function $p_Y: Y \to [0, \infty)$. Finally, let us denote $p_{Y|X}$ the conditional probability of $Y$ given $X$.\\
We get the mutual information between X and Y starting from its definition (equation \eqref{eqn:mutualInformation}).

\begin{align} 
I(X, Y) &= \mathbb{E}_{X,Y}\left[\log_2\left(\frac{p(Y | X)p(X)}{p(X)p(Y)}\right)\right] \notag \\
        &= \mathbb{E}_X\left[\int_{-\infty}^{\infty} p_Y(y | X)\log_2\left( \frac{p_Y(y | X)}{\sum\limits_{x' \in \Omega} p_{Y|X}(y | x')p_X(x')} \right) \diff y \right] \notag \\
        &= \sum\limits_{x \in \Omega} p_X(x) \int_{-\infty}^{\infty} p_{Y|X}(y | x)\log_2\left( \frac{p_{Y|X}(y | x)}{\sum\limits_{x' \in \Omega} p_{Y|X}(y | x') p_X(x')} \right) \diff y .\label{eqn:mutualInformationContinuousOutput}
\end{align}
\\
To get the results of the Section II of \cite{channelCoding}, we have to assume that the probability that a symbol occurs is uniformly distributed among $\Omega$, i.e. $p_X(x) = 1/n$. We can now express the mutual information as the following:
\begin{align}
I^*(X, Y) &= \frac{1}{n} \sum\limits_{x \in \Omega} \int_{-\infty}^{\infty} p_{Y|X}(y | x)\left[\log_2 n + \log_2\left( \frac{p_{Y|X}(y | x)}{\sum\limits_{x' \in \Omega} p_{Y|X}(y | x')} \right)\right] \diff y \notag\\
&=\log_2 n + \frac{1}{n} \sum\limits_{x \in \Omega} \int_{-\infty}^{\infty} p_{Y|X}(y | x)\log_2\left( \frac{p_{Y|X}(y | x)}{\sum\limits_{x' \in \Omega} p_{Y|X}(y | x')} \right) \diff y .
\label{eqn:mutualInformationUniformDistribution}
\end{align}
Recall that we send information over the AWGN channel. That means the noise is Gaussian with covariance matrix $\sigma^2I_m$. 
Let $W$ be the random variable representing the noise. It is defined as the difference between $Y$ and $X$, .i.e. $W = Y - X$.
We insert the probability density function of the noise in \eqref{eqn:mutualInformationUniformDistribution}:\\
\begin{align} 
I^*(X, Y) &= \log_2 n + \frac{1}{n} \sum\limits_{x \in \Omega} \int_{-\infty}^{\infty}
p_W(w) \log_2 \frac{e^{-\frac{|y - x|^2}{2\sigma^2}}}{\sum\limits_{x' \in \Omega} e^{-\frac{|y - x'|^2}{2\sigma^2}}} \diff y \notag \\ 
        &= \log_2 n - \frac{1}{n} \sum\limits_{x \in \Omega} \mathbb{E}_{W} \left[ \log_2 \sum\limits_{x' \in \Omega} e^{-\frac{|W + x - x'|^2 - |W|^2}{2\sigma^2}} \right] .
        \label{eqn:MIPaperUnger}
\end{align}
Now that we have the mathematical formula for the mutual information of the AWGN channel under a uniform input distribution, we will be able to plot the mutual information over the SNR.

\section{Results}

\begin{figure}[ht]
    \centering
    \subfloat[\centering MI over SNR of real constellations]{{\includegraphics[width=6cm]{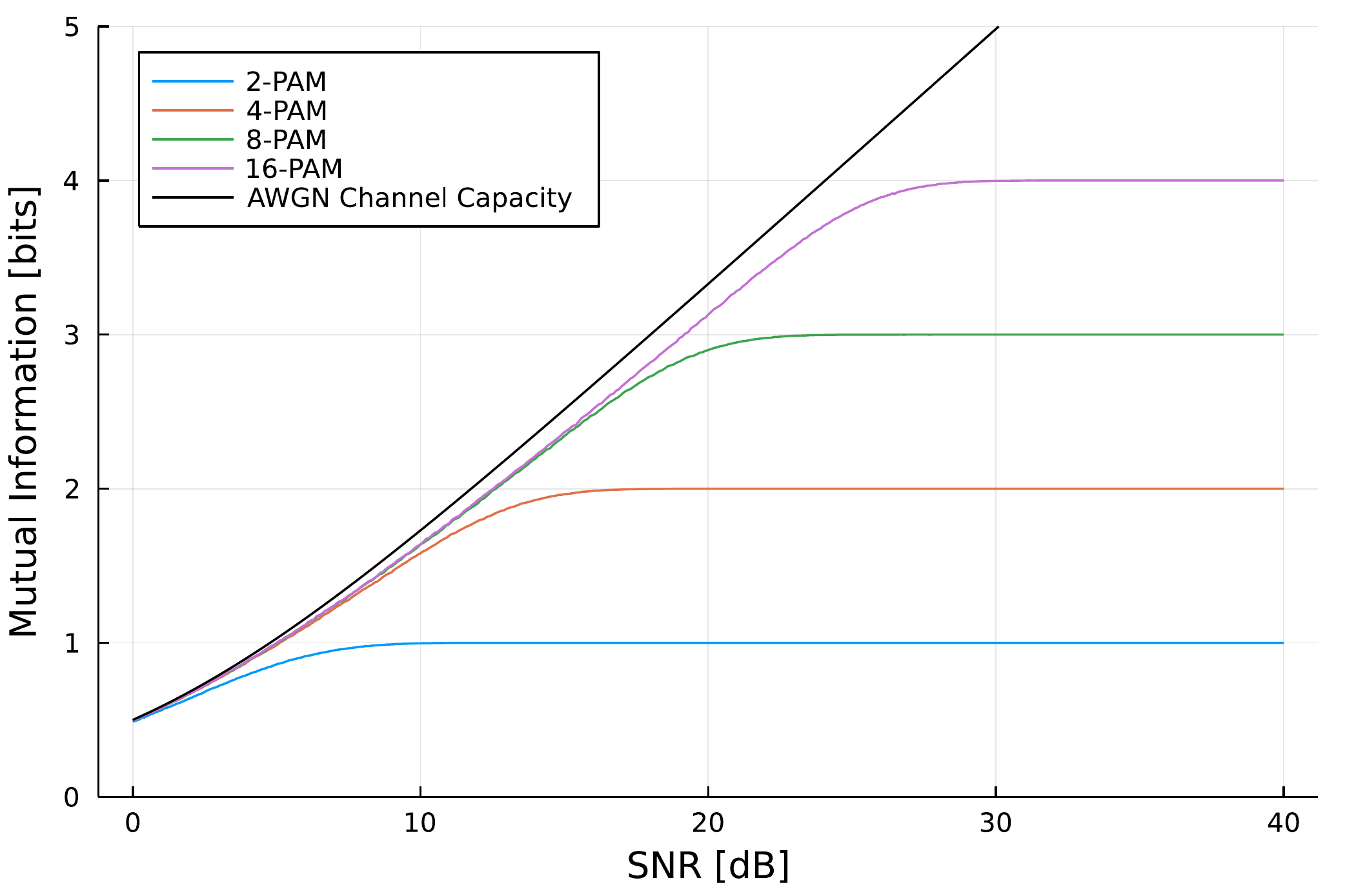} }}%
    \qquad
    \subfloat[\centering MI over SNR of complex constellations]{{\includegraphics[width=6cm]{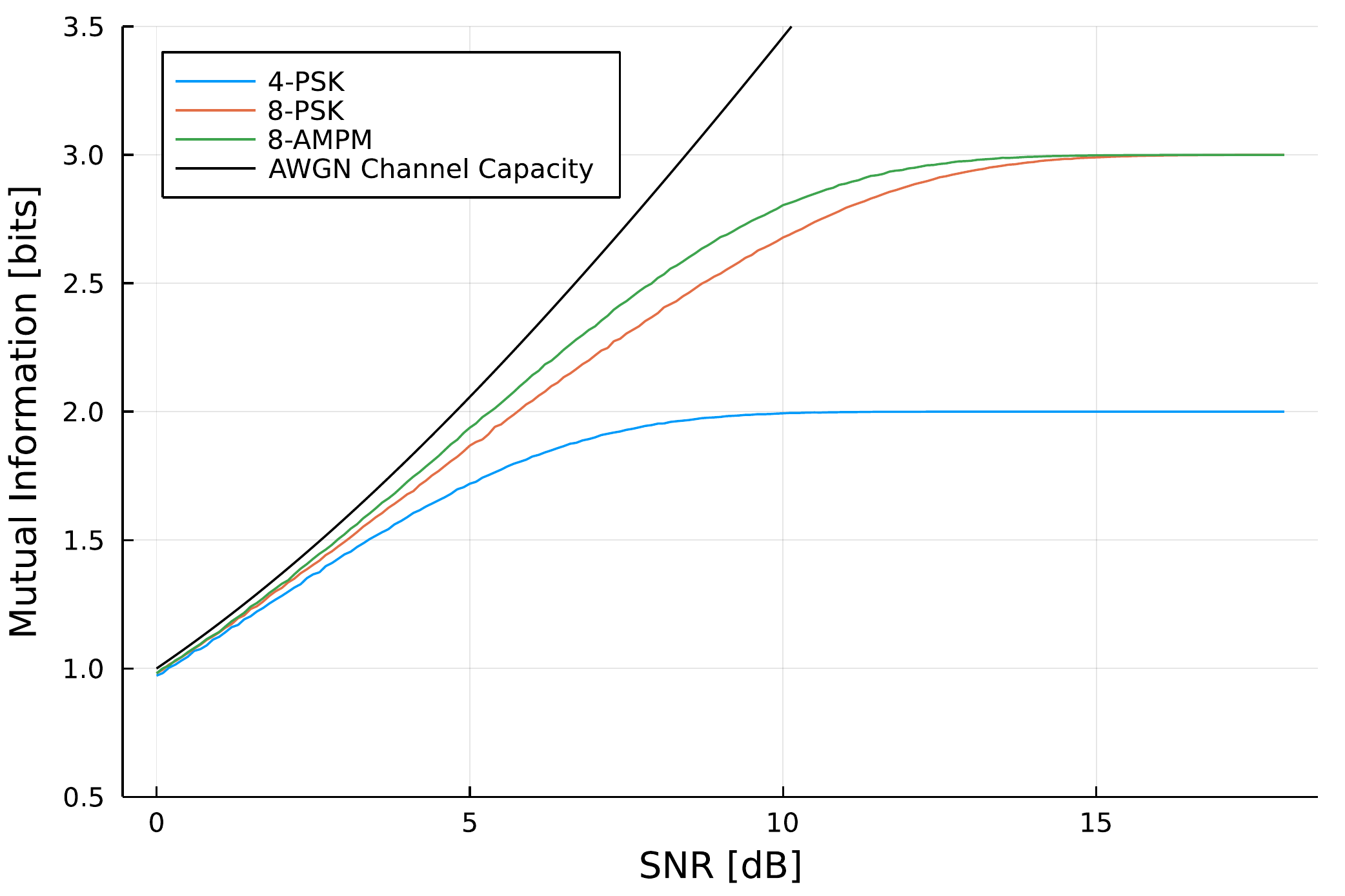} }}%
    \caption{Plots of capacity over SNR for different constellations}%
    \label{fig:capacityOverSNR}%
\end{figure}

\begin{figure}[ht]
    \centering
    \includegraphics[width=9cm]{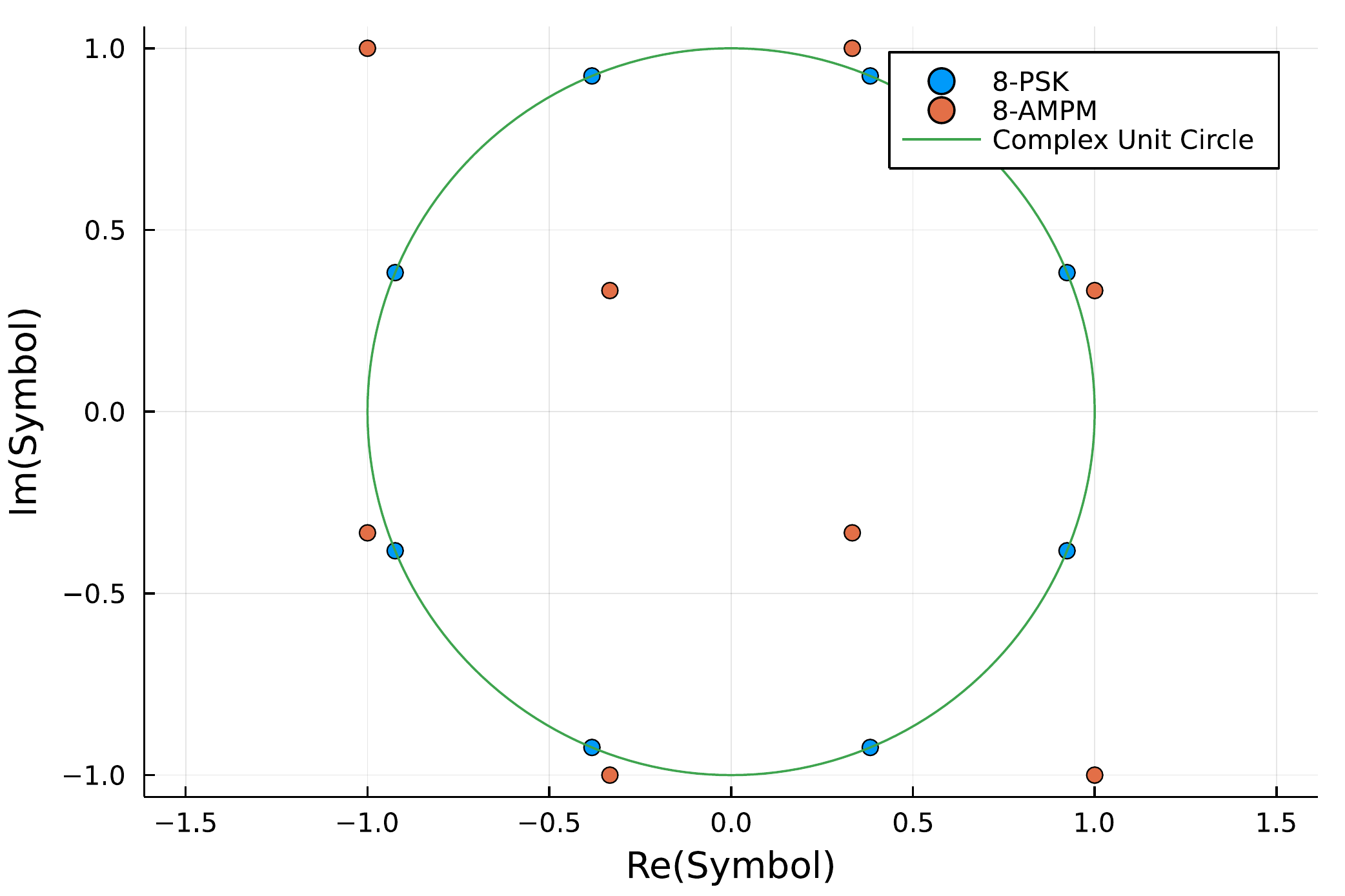}
    \caption{8-PSK and 8-AMPM constellations}
    \label{fig:8psk8ampmconstellations}
\end{figure}

To evaluate the expected value $E_W$ in \eqref{eqn:MIPaperUnger}, we use Monte Carlo simulations. More precisely, we generate a big vector of noise samples (in Figure \ref{fig:capacityOverSNR}, 100,000 noise samples per MI evaluation are used). We then compute the inner part of the expectation over this vector and average the values obtained. It leads to an approximation of the expected value with a precision that is sufficient for our needs.
\\\\
We want to plot the mutual information of $X$ and $Y$ over the SNR. We are essentially looking at how many bits per channel use we can send through the AWGN channel for a range of SNR. In our case, the relation between the SNR and $\sigma^2$ is
\[
\SNR = \frac{\mathcal{P}_{signal}}{\mathcal{P}_{noise}} =
  \begin{cases}
    \frac{\mathbb{E}[|X|^2]}{\sigma^2}       & \quad \text{if } m = 1\\
    \frac{\mathbb{E}[||X||^2]}{2\sigma^2}    & \quad \text{if } m = 2
  \end{cases} 
\]
\\
Figure \ref{fig:capacityOverSNR} shows the mutual information in bits over the SNR in decibel for different constellations having a uniform input distribution over the AWGN channel.\\
\\
We immediately observe that the mutual information of the 8-AMPM constellation is bigger for every SNR than the 8-PSK constellation, even though they have the same number of symbols in their alphabet.\\
This is due to the fact that the minimum Euclidean distance between each symbol is greater for 8-AMPM (see Figure \ref{fig:8psk8ampmconstellations}) than for 8-PSK. Indeed, as the noise is Gaussian, this leads to a lower probability that an output gets interpreted as the wrong symbol.\\

\section{Generalisation to a Nonuniform Input Distribution }

In \eqref{eqn:mutualInformationUniformDistribution}, we have an expression for the mutual information for constellations that have a uniform input distribution. We want to generalise it for any input distribution. To accomplish this, we start from \eqref{eqn:mutualInformationContinuousOutput} and replace the PDF of the noise, but this time we keep the input distribution in the equation:\\
\begin{align}
    I(X, Y) &= \sum\limits_{x \in \Omega}p_X(x) \int_{-\infty}^{\infty} - p_{Y|X}(y|x) \log_2 \frac{\sum\limits_{x'\in X}p_X(x')e^{-\frac{(y-x')^2}{2\sigma^2}}}{e^{-\frac{(y-x)^2}{2\sigma^2}}} \diff y \notag \\
    &= - \sum\limits_{x \in \Omega}p_X(x) \mathbb{E}_W \left[ \log_2 \sum\limits_{x' \in \Omega}p_X(x') e^{-\frac{|W + x - x'|^2 - |W|^2}{2\sigma^2}} \right].
\end{align}
\\
Let's take a look at what happens for a 2-PAM (the constellation is $\{-1;1\}$) with nonuniform probability distribution.
\begin{figure}[ht]
    \centering
    \includegraphics[width=9cm]{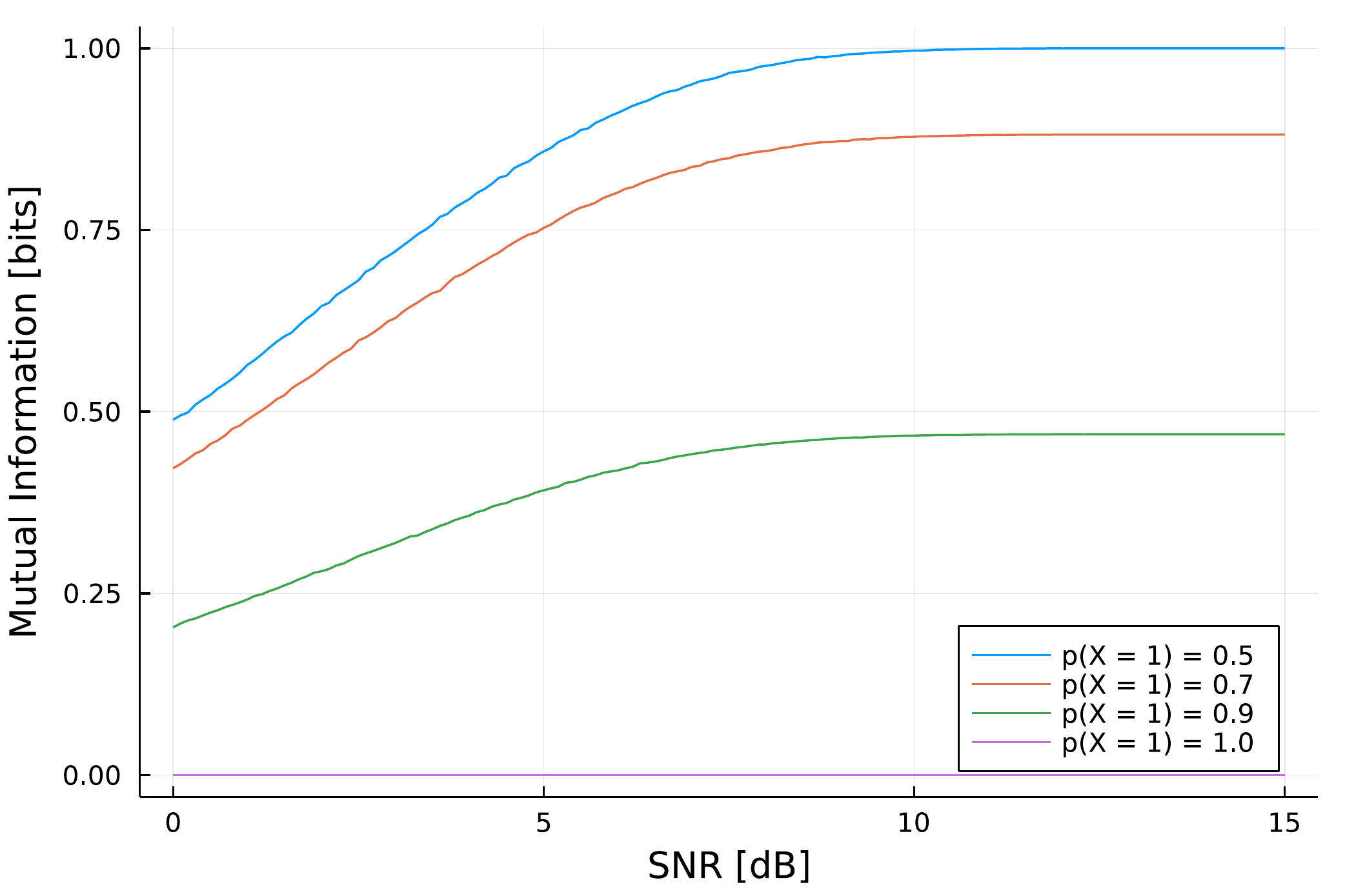}
    \caption{MI of a 2-PAM with different input distributions over SNR}
    \label{fig:2pamUnbalanced}
\end{figure}
\\
We clearly see in Figure \ref{fig:2pamUnbalanced} that the mutual information achieved by the uniform distribution is better. Especially, at high SNR, this is because the mutual information measures how much we can know about $X$ by observing $Y$. But if the information from $X$ reduces, i.e. its entropy reduces, the mutual information is also reduced. Recall the definition of the mutual information $I(X,Y) = H(X) - H(X|Y)$. The mutual information of $X$ clearly appears in this definition, which explains why the mutual information reduces with this nonuniform input distribution. In section 4, we will try to exploit the possibility of choosing the input probability distribution in order to maximise the mutual information.
\\\\
It was convenient to use Monte Carlo simulations to evaluate the mutual information. However, if we want to obtain the results faster and with a better precision we need to quantize the output of the channel. It will also help us to improve iteratively the mutual information later. In the next chapter, we are going to look at the mutual information obtained over an output-quantized AWGN channel.

%% file: Chapters/3MI_Discrete.tex
%%%%%%%%%%%%%%%%%%%%%%%%%%%%%%%%%%%%%%%%%%%%%%%%%%%%%%%%%%%%%%%%%%%%%%%%
\chapter{Mutual Information for a Discrete-Output Channel}
%%%%%%%%%%%%%%%%%%%%%%%%%%%%%%%%%%%%%%%%%%%%%%%%%%%%%%%%%%%%%%%%%%%%%%%%

\begin{figure}[ht]
    \centering
    \includegraphics[width=9cm]{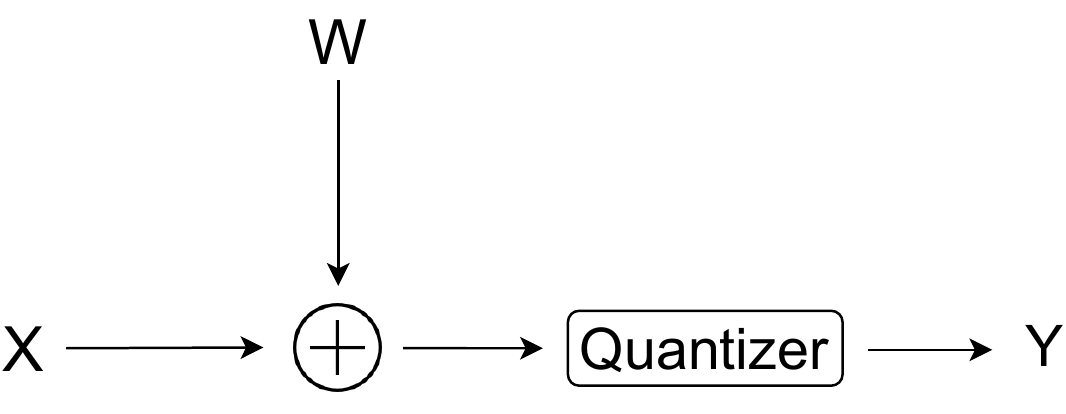}
    \caption{Diagram of the quantized AWGN channel}
    \label{fig:DiagramAWGNQuantized}
\end{figure}

In this section, we will estimate the mutual information of an AWGN channel for a fixed finite constellation, a fixed input distribution and a quantized output.
\\\\
Suppose that the input signal $X$ and the output signal $Y$ are contained in a vector space $V$ of dimension $m$. The real AWGN channel is represented by $m=1$ and $V = \real$. The complex AWGN channel is represented by $m=2$ and $V = \mathbb{R}^2$. We assume $X$ takes values in $\Omega \subset V$. For the moment, the output of the channel is continuous as the noise $W$ is continuous. We will now suppose that the output $Y$ takes values in a finite set $\mathcal{Q} \subset V$. Thus, it is a discrete random variable. Figure \ref{fig:DiagramAWGNQuantized} summarises the new situation.

\section{Formula for the MI}

The formula for the mutual information obtained over a quantized AWGN channel is very similar to the one with continuous output (equation \eqref{eqn:mutualInformationContinuousOutput}). The difference is that in the discrete case, $\mathbb{E}_Y[\cdot]$ will not result in an integral, but in a sum over all $y \in \mathcal{Q}$:
\\
\begin{align} 
I(X, Y) &= \mathbb{E}_{X,Y}\left[\log_2\left(\frac{p(Y | X)p(X)}{p(X)p(Y)}\right)\right] \notag \\
        &= \sum\limits_{x \in \Omega} p_X(x) \sum\limits_{y \in \mathcal{Q}} p_{Y|X}(y | x)\log_2\left( \frac{p_{Y|X}(y | x)}{\sum\limits_{x' \in \Omega} p_{Y|X}(y | x') p_X(x')} \right)  .\label{eqn:mutualInformationDiscreteOutput}
\end{align}
\\
Now that $Y$ takes values in a finite set, if we restrict ourselves to PAM and QAM constellations, we can compute analytically $p(y|x) \forall x \in \Omega, y \in \mathcal{Q}$.
\\
We have the formula for computing the mutual information, but it is still unclear how the quantizer should be chosen.

\section{The Quantizer}

To make the output discrete, we have to quantize it.
The quantizer maps $y \in V$ to $q \in \mathcal{Q}$ such that the distance between $q$ and $y$ is minimised, i.e.
\[
|| q - y || \leq || q' - y ||, \forall q' \in \mathcal{Q} ,
\]
where $||\cdot||$ is the Euclidean norm.
\\
Now, we must study how to choose $\mathcal{Q}$ in order to maximise the mutual information without adding too much computational complexity. In the case of QAM constellations, we will use cross points of a grid centred around the mean.

\subsection{Choosing the bounds of the quantizer}

\begin{figure}[ht]
    \centering
    \includegraphics[width=12cm]{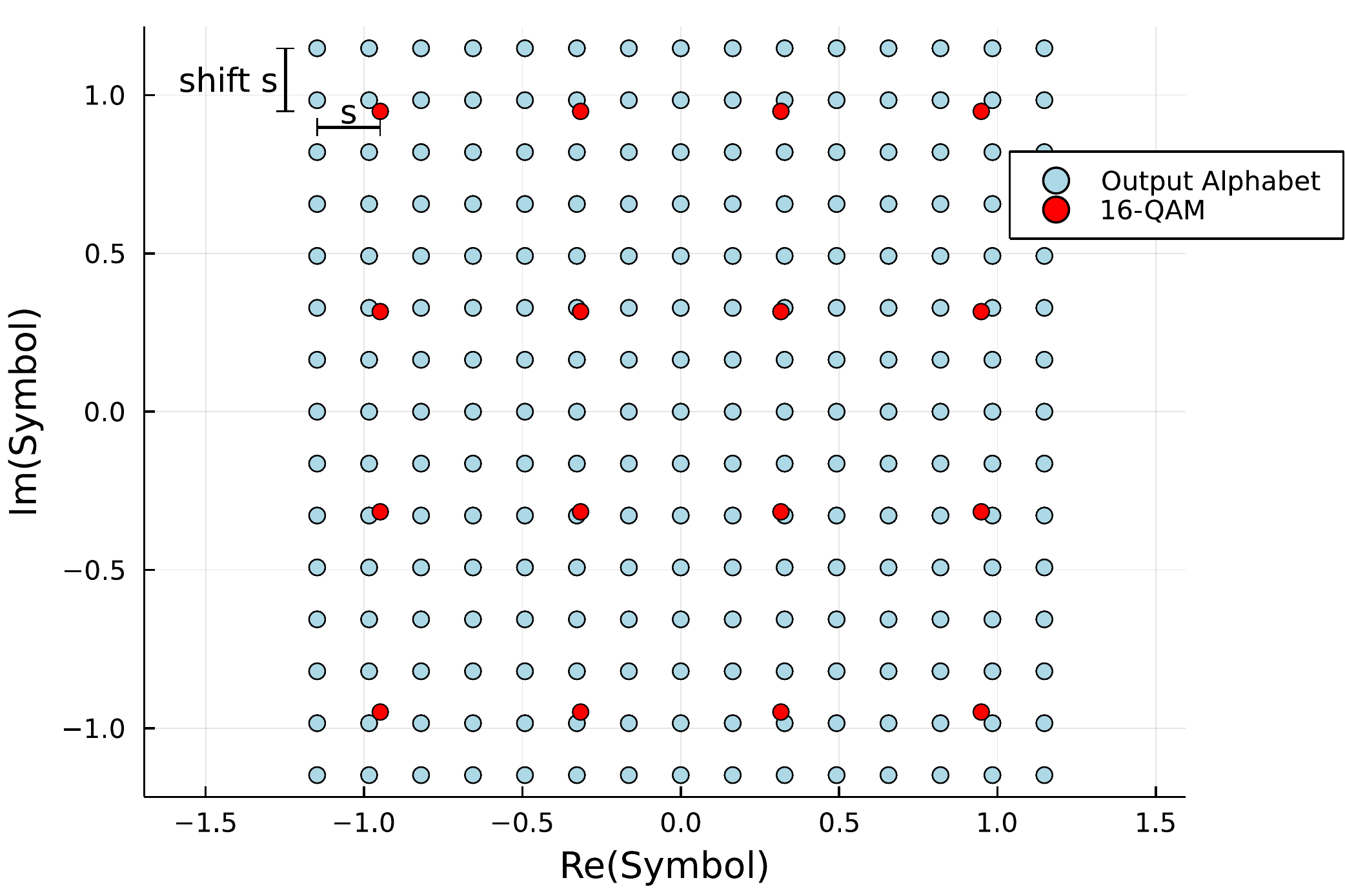}
    \caption{Example of output alphabet for a 16-QAM}
    \label{fig:16QAMQuantizerExample}
\end{figure}

\begin{figure*}
    \centering
    \begin{subfigure}[b]{0.475\textwidth}
        \centering
        \includegraphics[width=\textwidth]{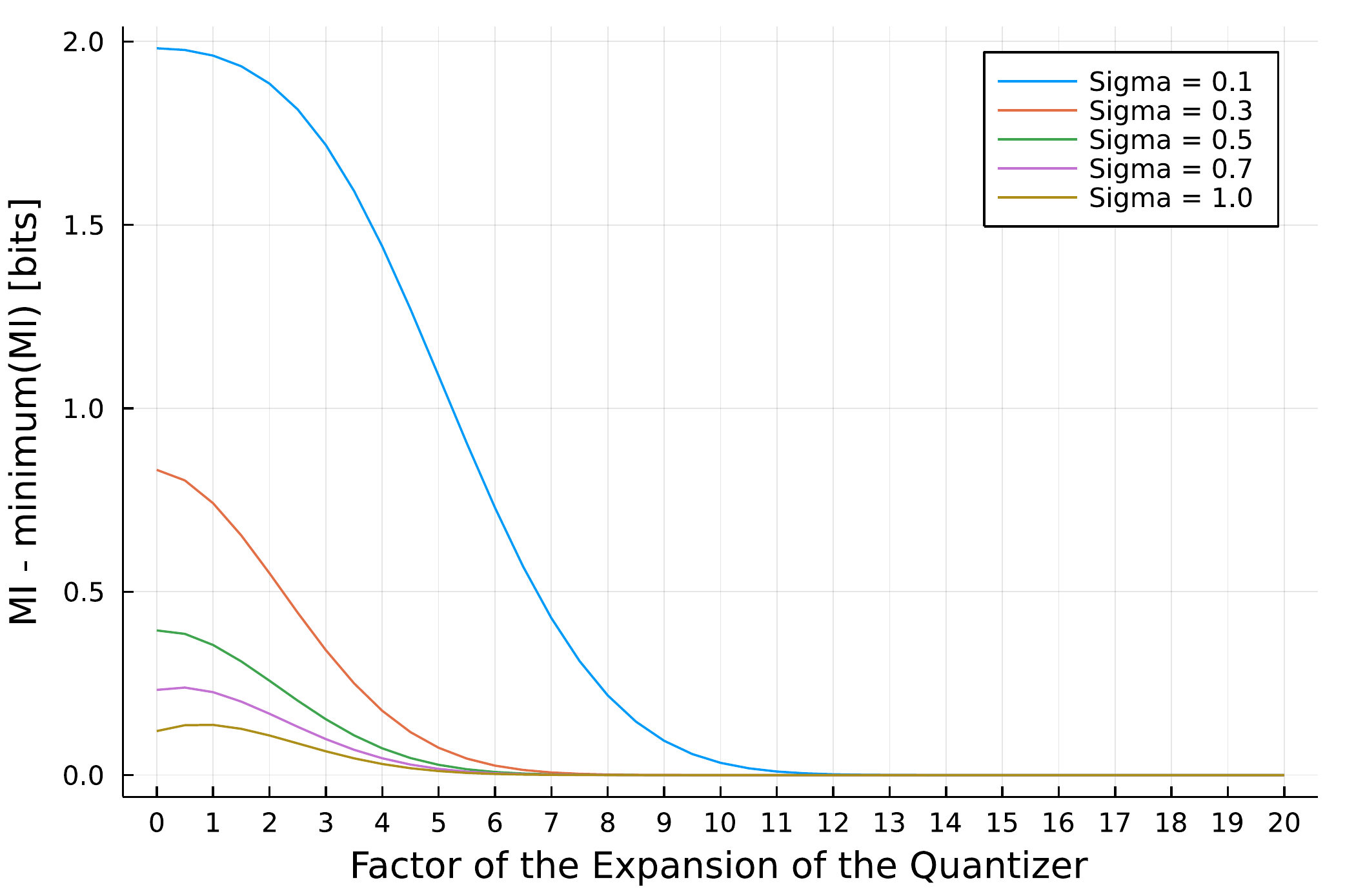}
        \caption[]%
        {{\small Quantizer using 2 bits/dimension}}    
        \label{fig:quantizer shift 2 bits}
    \end{subfigure}
    \hfill
    \begin{subfigure}[b]{0.475\textwidth}  
        \centering 
        \includegraphics[width=\textwidth]{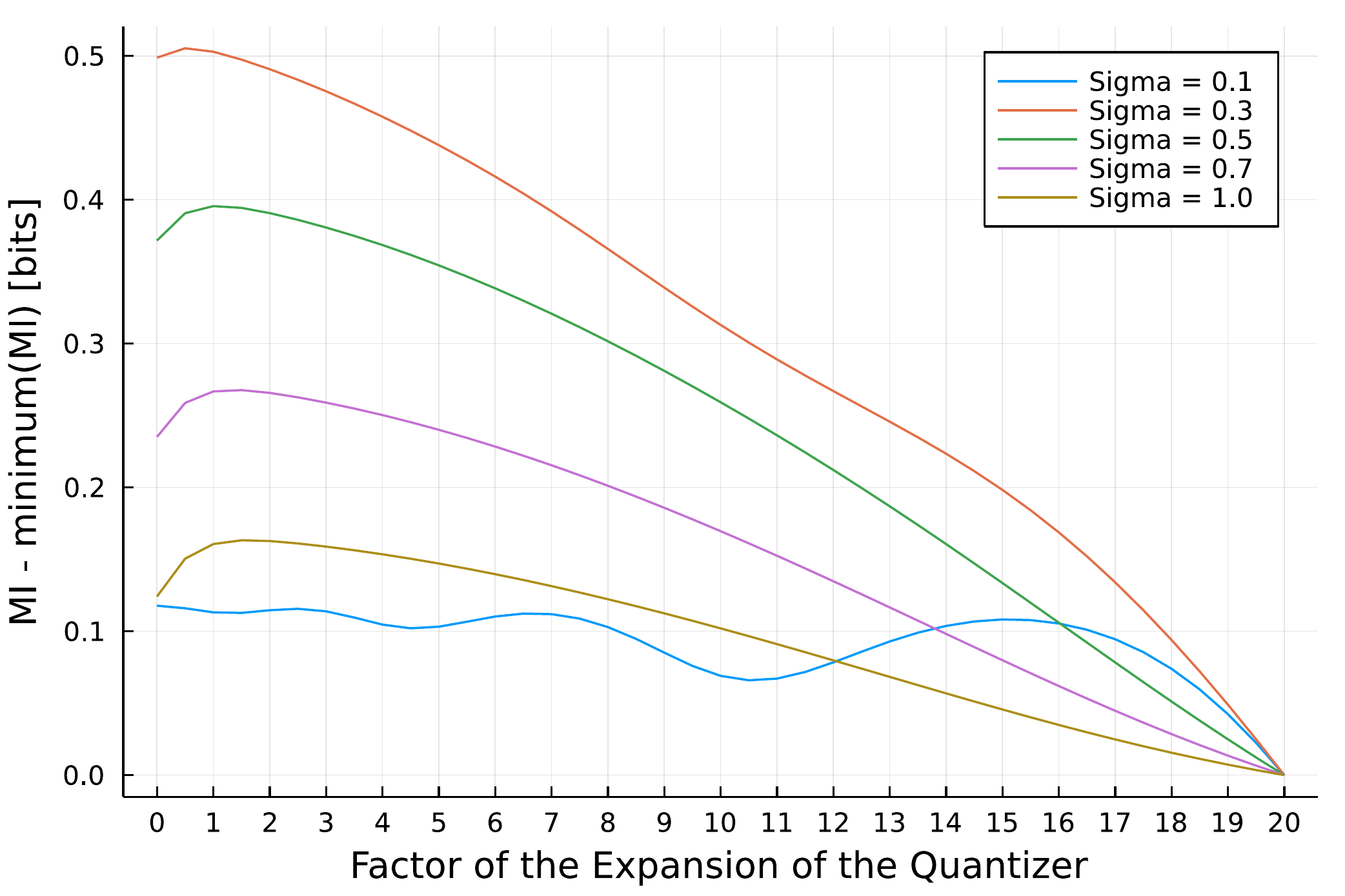}
        \caption[]%
        {{\small Quantizer using 4 bits/dimension}}    
        \label{fig:quantizer shift 4 bits}
    \end{subfigure}
    \vskip\baselineskip
    \begin{subfigure}[b]{0.475\textwidth}   
        \centering 
        \includegraphics[width=\textwidth]{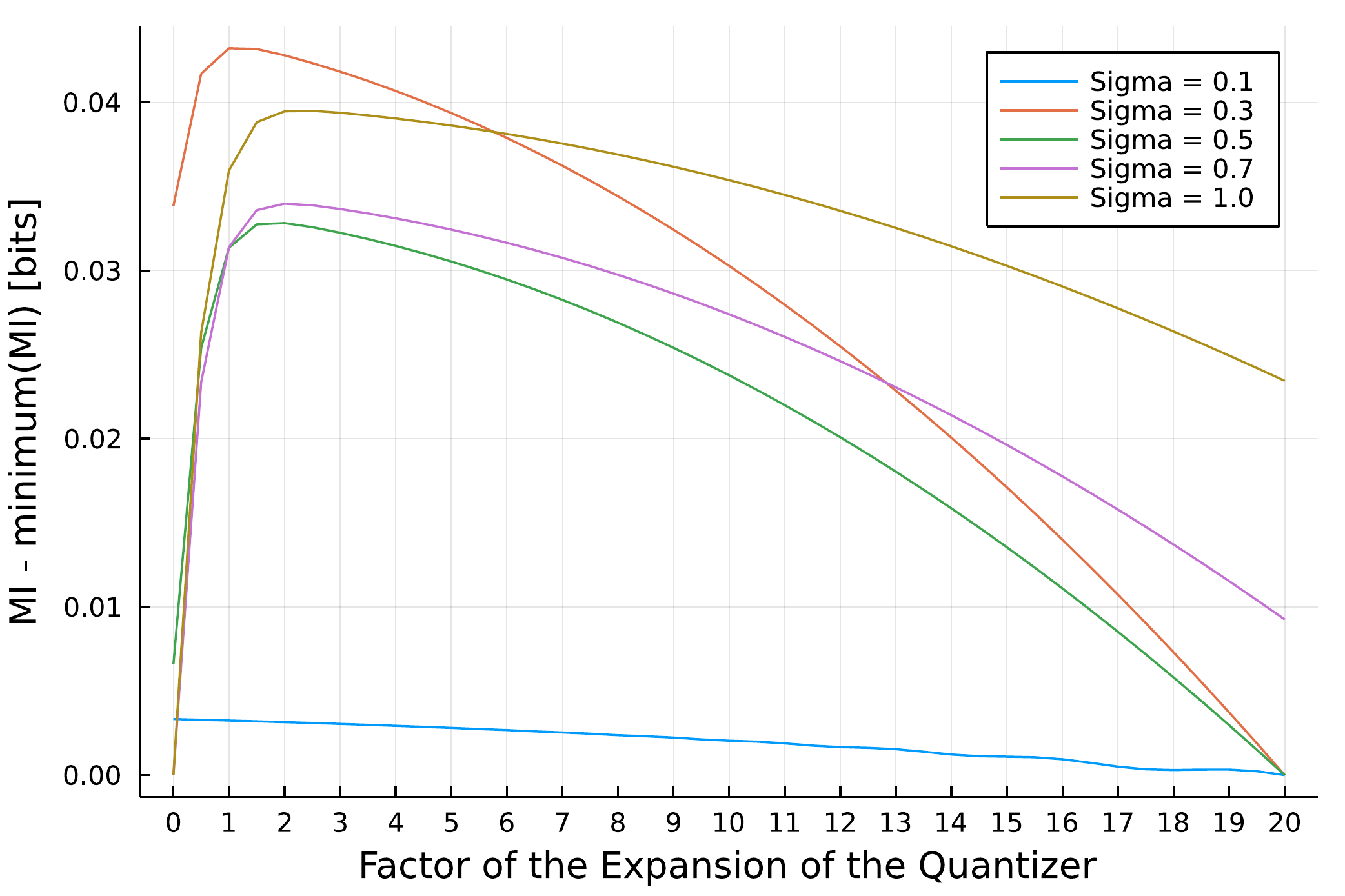}
        \caption[]%
        {{\small Quantizer using 6 bits/dimension}}    
        \label{fig:quantizer shift 6 bits}
    \end{subfigure}
    \hfill
    \begin{subfigure}[b]{0.475\textwidth}   
        \centering 
        \includegraphics[width=\textwidth]{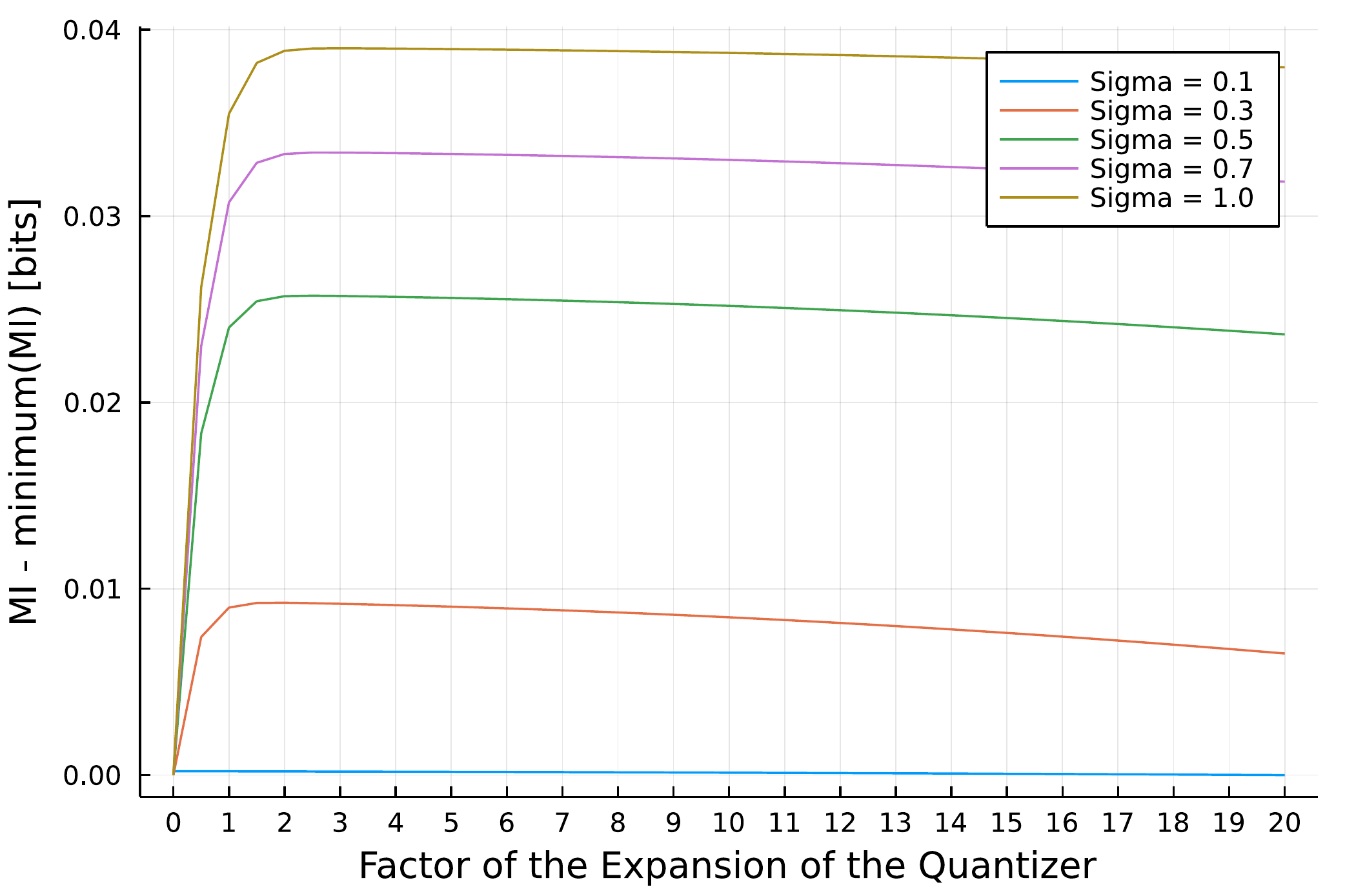}
        \caption[]%
        {{\small Quantizer using 8 bits/dimension}}    
        \label{fig:quantizer shift 8 bits}
    \end{subfigure}
    \caption{The difference of MI versus $s / \sigma$ for different quantizers}
    \label{fig:mi for different quantizers}
\end{figure*}

We choose to use a uniform quantizer in each dimension for the output alphabet. We have to ask ourselves what are the boundaries of this quantizer. Should they be bigger than the outer points of the constellation or should they be located at the same position?
\\
Figure \ref{fig:16QAMQuantizerExample} shows an example of an output alphabet using 4 bits of quantization per dimension. Changing the shift $s$ of the outer points will result in a smaller resolution, but it will also match more closely the output signals that ended out of the bounds of the constellation.
\\
To determine the right value for the shift, we generate the mutual information over multiples of the standard deviation of the noise. Indeed, we will use a shift that depends on $\sigma$. We also generate the mutual information for different number of bits of quantization.
\\
Figure \ref{fig:mi for different quantizers} shows the mutual information achieved by a 16-QAM constellation over a uniformly distributed input for different number of bits of quantization, different standard deviations $\sigma$ of the noise and different shifts. The mutual information achieved when $\sigma$ is high is significantly different then when $\sigma$ is low. To be able to make comparisons on the same graph, we plot the mutual information minus the minimum mutual information calculated for each curve. This means that each curve must have its lowest value equal to zero.
\\
The horizontal axis shows factors of $\sigma$, i.e. $s = \text{value on the horizontal axis} \cdot \sigma$.
\\
On the plots, we see that the improvement in mutual information obtained won't be noticeable when plotting the mutual information. However, as we will later want to get as close as possible to the capacity, it will be important to be careful about the choice of the quantizer.
\\
For a 16-QAM, a quantizer using two bits per dimension is clearly not precise enough. Thus, we are not going to take our decision for the right shift value based on Figure~\ref{fig:quantizer shift 2 bits}. The curves obtained in the other figures have more or less the same shape. First, the mutual information increases rapidly, and then it decreases slowly as the resolution of the quantizer gets worse.
\\
We see that using a shift of two times the standard deviation, $s = 2 \cdot \sigma$, gives us a mutual information close to the maximum that we can achieve in the Figures~\ref{fig:quantizer shift 4 bits}, \ref{fig:quantizer shift 6 bits} and \ref{fig:quantizer shift 8 bits}. Thus we are going to choose this shift if in the next section we estimate that a quantizer using between four and eight bits per dimension is reasonable.

\subsection{Choosing the resolution of the quantizer}

We have to find a trade off between using more bits of quantization, and therefore achieving a bigger mutual information (as the quantization noise reduces), and the computation time to evaluate the mutual information. To find the number of bits that we will use, we generate the mutual information of a very noisy channel. We choose the standard deviation of the noise to be $\sigma = 1$. The Table~\ref{table:16 QAM MI bits of quantization} shows the mean of one thousand evaluations of the mutual information. It was obtained for a 16-QAM constellation with shift $s = 2 \cdot \sigma = 2$.
\\
\begin{table}[ht]
\centering
\begin{tabular}{ |p{4cm}||p{3cm}|p{3cm}|p{3cm}| }
 \hline
 Bits per dimension & mean(MI) & Time Taken\\
 \hline
 2 & 0.49203 & Very fast \\
 4 & 0.57735 & Fast \\
 5 & 0.58168 & Medium \\
 6 & 0.58277 & Medium to slow\\
 7 & 0.58304 & Very slow\\
 \hline
\end{tabular}
\caption{MI of a 16-QAM with different bits of quantization}
\label{table:16 QAM MI bits of quantization}
\end{table}
\\
Note that the sample variance of the mutual information is in each case negligible (less than $10^{-27})$. In the column "Time Taken", "Medium" means that it took less than ten seconds and "Very slow" means that it took between one and two minutes\footnote{The CPU used for the calculations is Intel(R) Core(TM) i7-8650U CPU @ 1.90GHz with 16GB RAM.}.
\\
For our needs, we will use five bits per dimension to quantize a 16-QAM and six bits per dimension for a 64-QAM. Concerning the PAM constellations, we can use a high number of bits of quantization without worrying about the time complexity of computing the mutual information. In practice, we will use eight or nine bits of quantization for PAM constellations.

\section{Results}

\begin{figure}[ht]
    \centering
    \includegraphics[width=9cm]{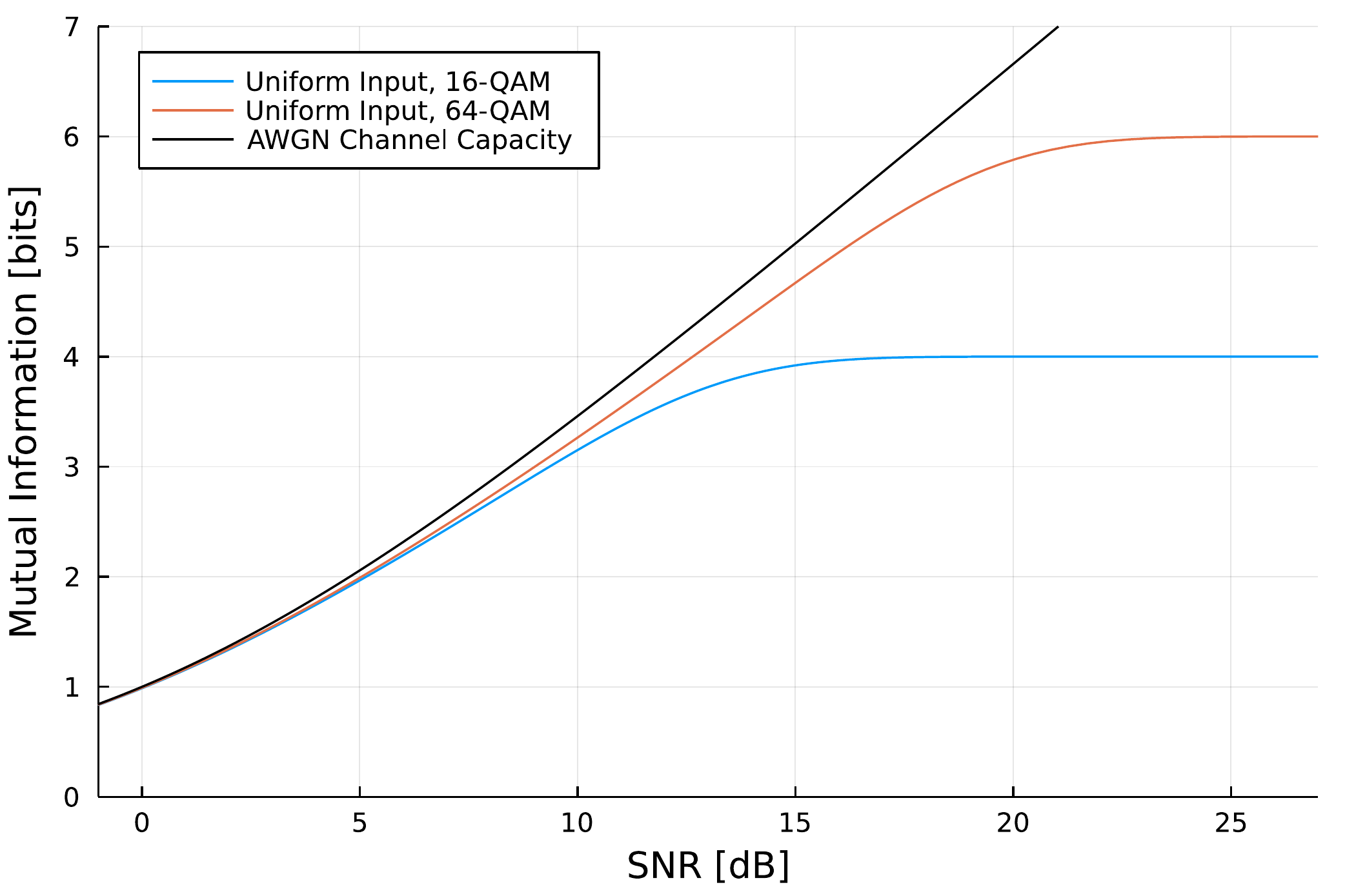}
    \caption{MI over SNR for 16-QAM and 64-QAM constellations with a uniform input distribution}
    \label{fig:MI Discrete Output Uniform}
\end{figure}

Figure \ref{fig:MI Discrete Output Uniform} shows the mutual information over the SNR. The input is uniformly distributed. As determined above, the quantizer uses a shift of twice the standard deviation of the noise, five bits of quantization for the 16-QAM and six bits of quantization for the 64-QAM.
\\
We notice that the curves are smoother than the curves in Figure~\ref{fig:capacityOverSNR}. Indeed, we are getting a more precise result as the conditional probability mass function of the channel has been replaced by the computation of $p(Y|X)$ which we can determine analytically instead of using the Monte Carlo simulations.
\\\\
On the Figure~\ref{fig:MI Discrete Output Uniform}, we notice that the mutual information achieved by both constellations do not match closely the AWGN channel capacity. But as this is a fundamental limit, we know that it can be approached arbitrarily closely. Shaping the constellation can help achieving a better mutual information for a given SNR. We can do probabilistic shaping or geometric shaping. On the one hand, probabilistic shaping refers to the act of changing the input probability distribution to improve the mutual information achieved. On the other hand, geometric shaping is a method in which we change the symbols of the constellation without changing the probability of sending them. It is easier to build electrical systems that use QAM constellations than it is for arbitrary constellations. It is also possible to compute analytically $p(Y|X)$ for a QAM constellation. Additionally, for other constellations, it is usually a computationally complex task to get a closed-form formula. Therefore, we choose to study probabilistic shaping in the rest of the report.

%% file: Chapters/4MB_Distribution.tex
\chapter{The Maxwell--Boltzmann Distribution}

\begin{figure*}[ht]
    \centering
    \begin{subfigure}[b]{0.475\textwidth}
        \centering
        \includegraphics[width=\textwidth]{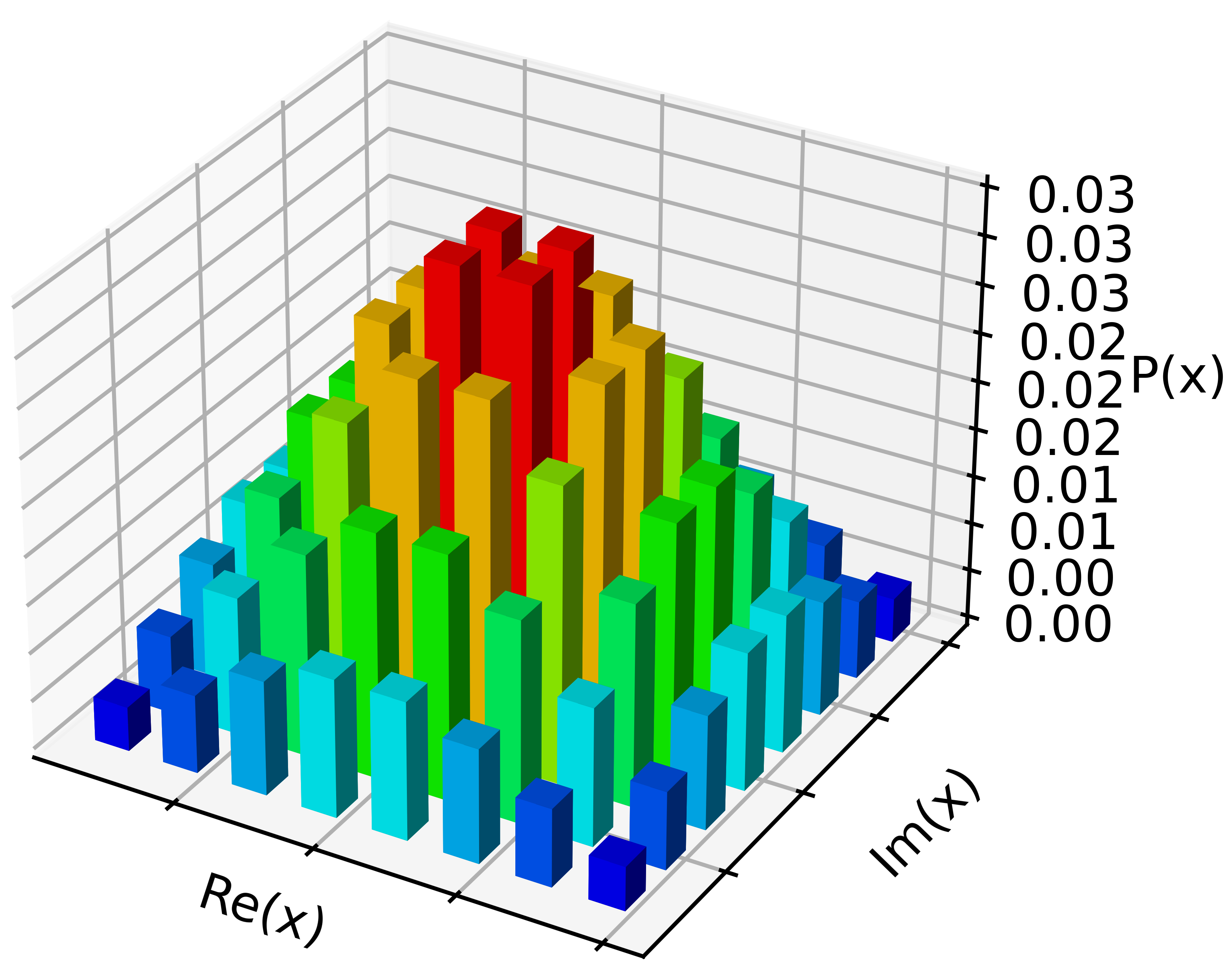}
        \caption[]%
        {{\small MB with $\lambda = 1$}}    
        \label{fig:MB lambda 1 probability distribution}
    \end{subfigure}
    \hfill
    \begin{subfigure}[b]{0.475\textwidth}  
        \centering 
        \includegraphics[width=\textwidth]{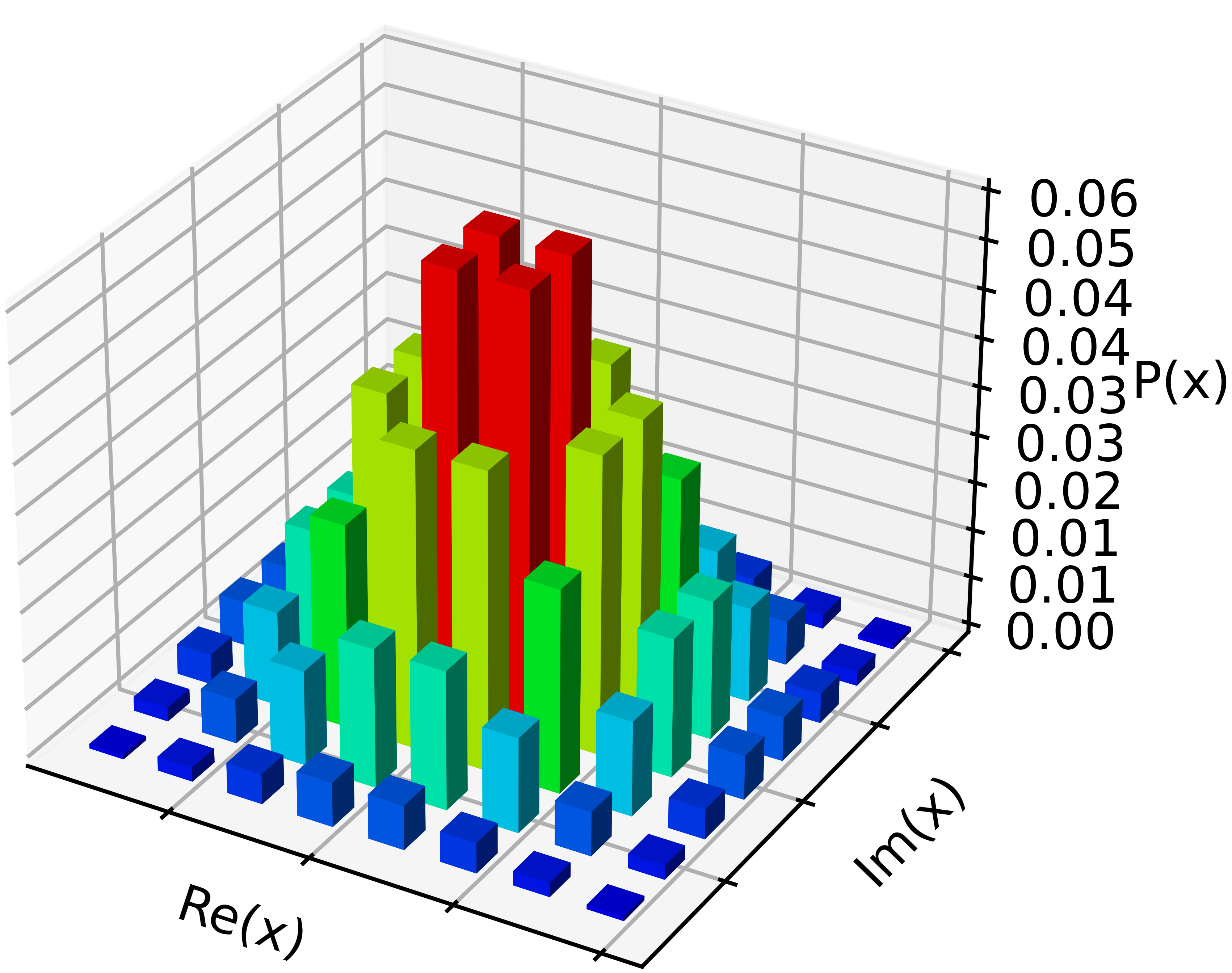}
        \caption[]%
        {{\small MB with $\lambda = 2$}}    
        \label{fig:MB lambda 2 probability distribution}
    \end{subfigure}
    \caption{Examples of the MB distribution for a 64-QAM}
    \label{fig:Example MB probability distribution 3d view}
\end{figure*}

\begin{figure}[ht]
    \centering
    % small image is scaled to 32%
    \includegraphics[width=8cm]{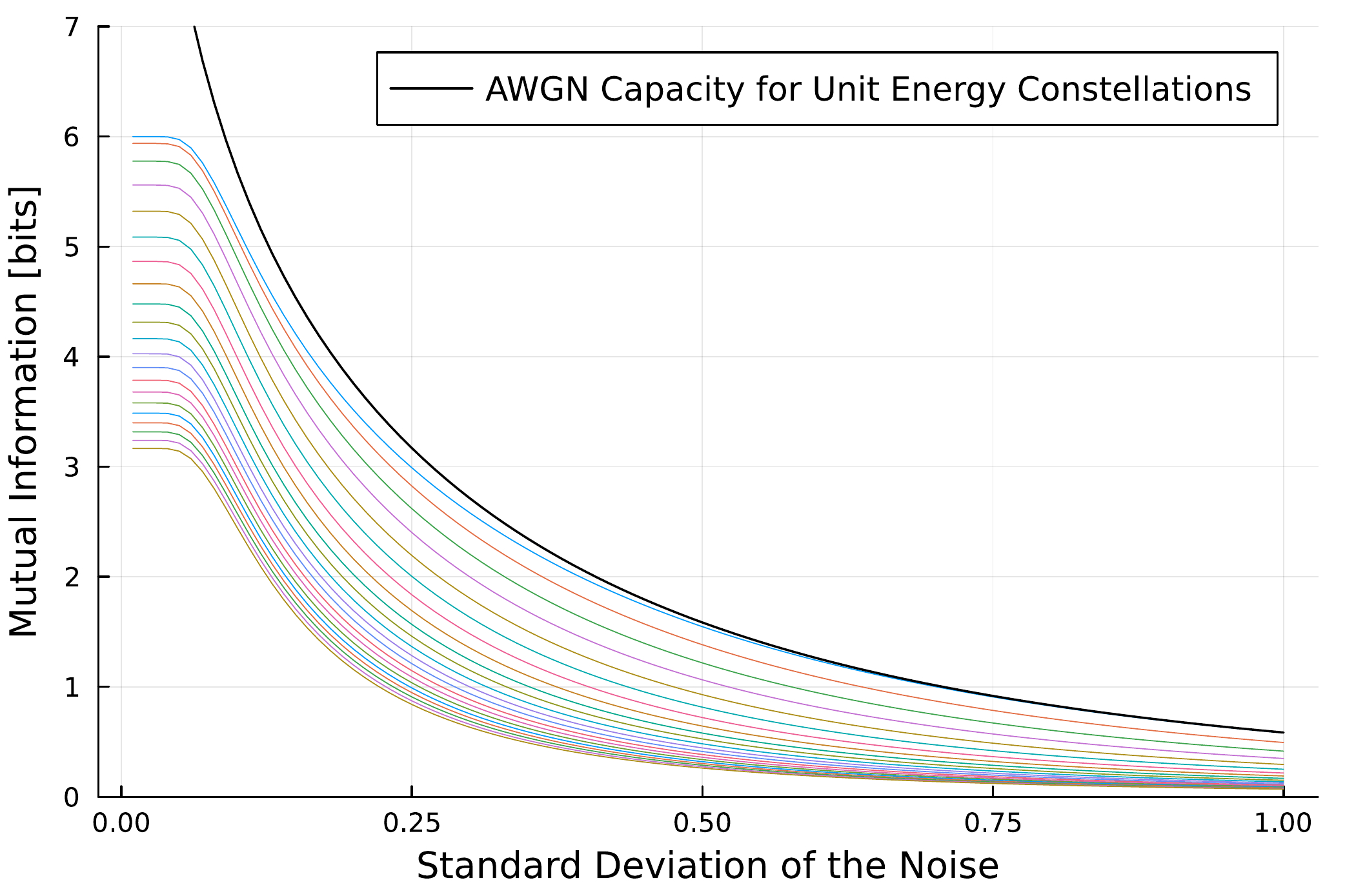}
    \caption{MI over $\sigma$, 64-QAM, the thin curves are generated by the MB distribution}
    \label{fig:MB multiple lambda not scaled}
\end{figure}

\begin{figure}[ht]
    \centering
    % small image is scaled to 32%
    \includegraphics[width=12cm]{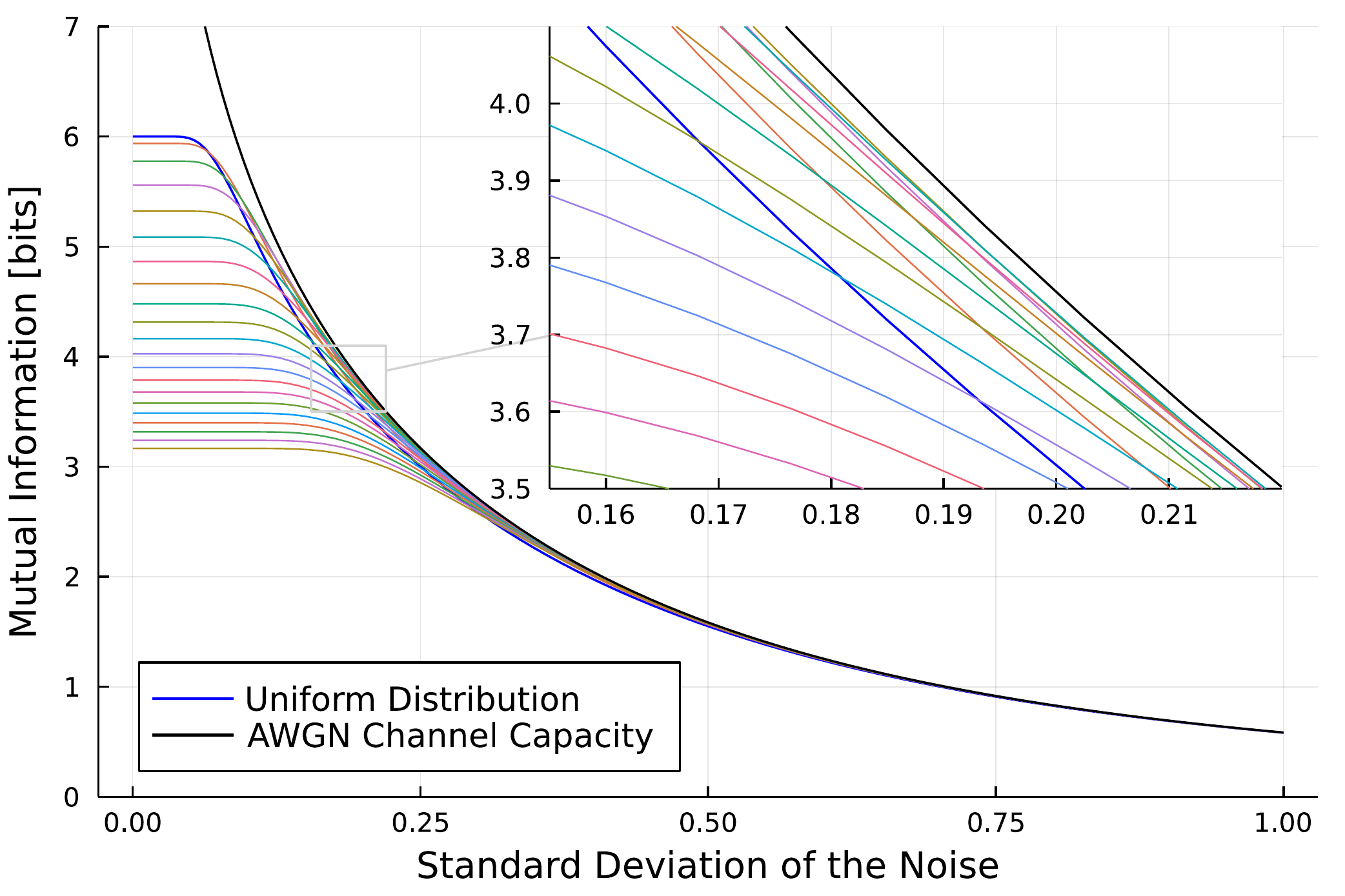}
    \caption{MI over $\sigma$, 64-QAM, the thin curves are generated by the MB distribution, the constellation has unit energy}
    \label{fig:MB multiple lambda}
\end{figure}

\begin{figure*}[ht]
    \centering
    \begin{subfigure}[b]{0.475\textwidth}
        \centering
        \includegraphics[width=\textwidth]{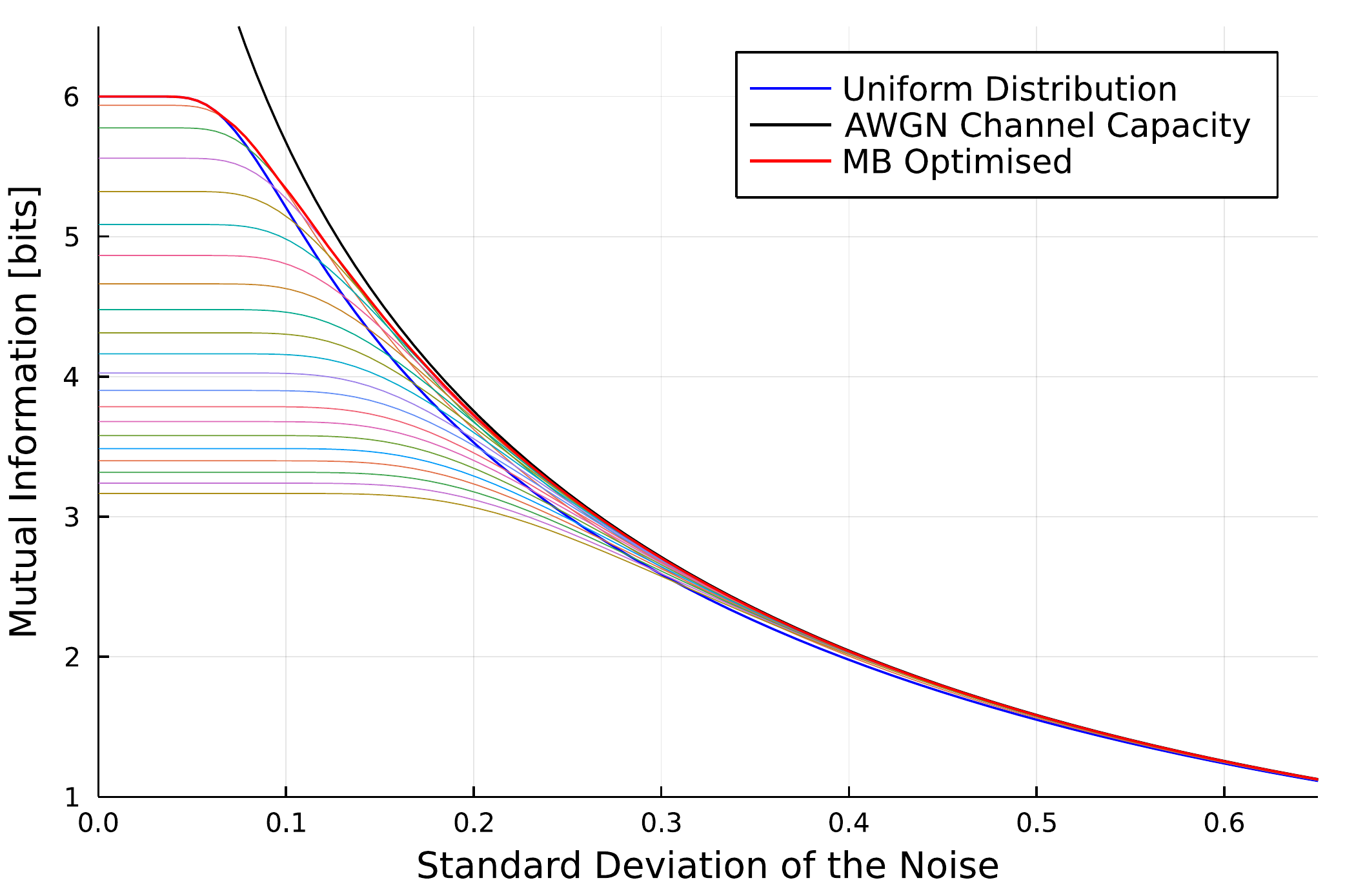}
        \caption[]%
        {{\small MB compared to Uniform over $\sigma$}}    
        \label{fig:Comparison MB UNI QAM64 over sigma}
    \end{subfigure}
    \hfill
    \begin{subfigure}[b]{0.475\textwidth}  
        \centering 
        \includegraphics[width=\textwidth]{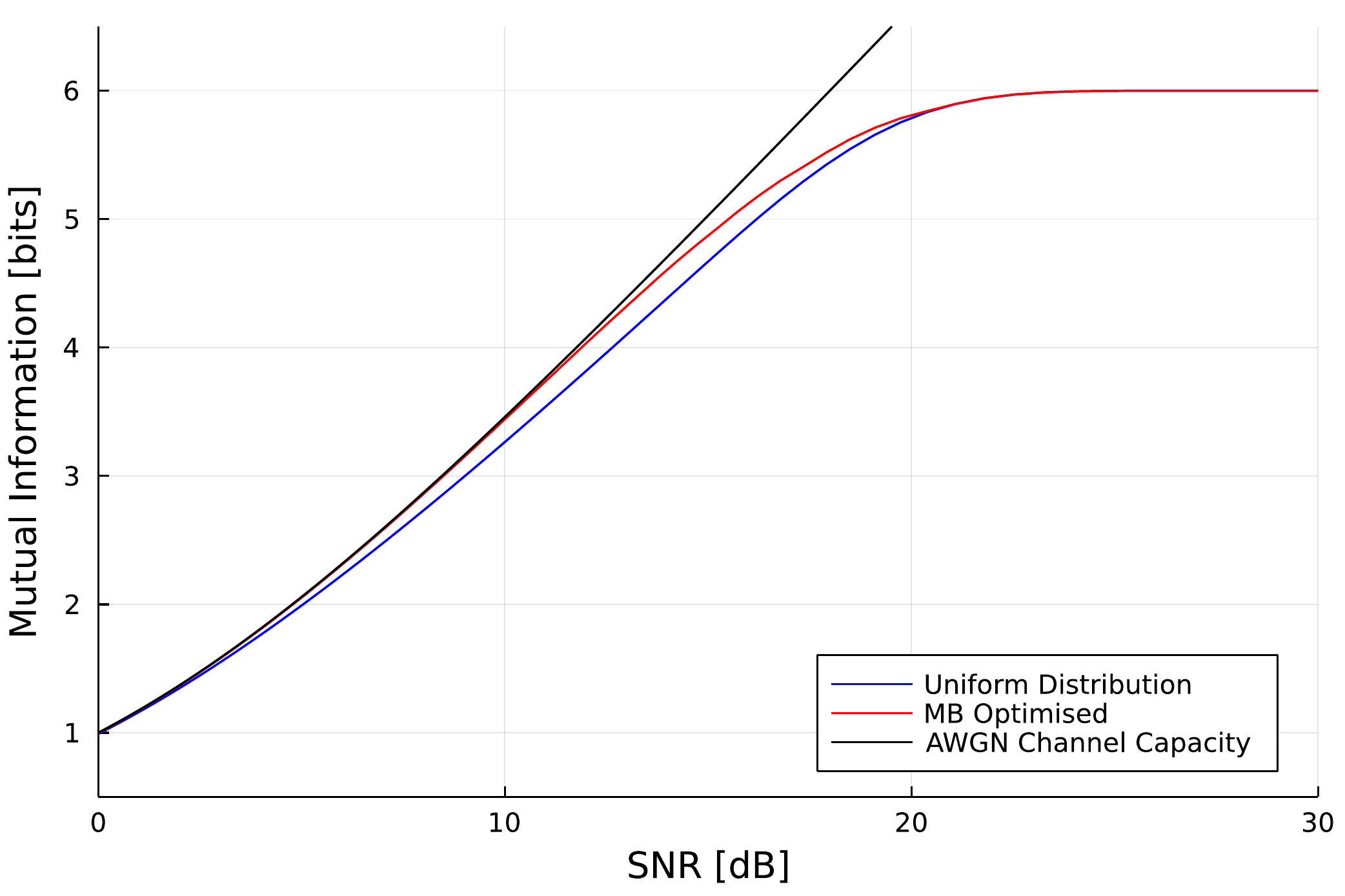}
        \caption[]%
        {{\small MB compared to Uniform over SNR}}    
        \label{fig:Comparison MB UNI QAM64 over SNR}
    \end{subfigure}
    \caption{Comparison between the MB and the uniform distribution for a 64-QAM}
    \label{fig:Comparison MB UNI QAM64}
\end{figure*}

If we want to get close to the AWGN channel capacity, we could arbitrarily increase the energy of the constellation. In practice, we would increase the inter-symbol distance of the constellation. Clearly, if we allow ourselves to use a constellation having a lot more energy, the probability of an error occurring while decoding decreases.\\
In practice, we are always limited by the energy of the constellation. For example, when a device sends data over the WiFi, the energy of the signal is restricted by the law to avoid interference with other frequencies. An other example is in fibre optic communications, using a signal having too much energy will damage the fibre and will sometimes result in the effect of fibre fuse.\\
\\
As explained at the end of Section 3, we are going to use probabilistic shaping to improve the mutual information achieved. \\
We know that we are limited by the energy of the constellation. We also know that the mutual information is given by a difference between the entropy of the input $X$ and the conditional entropy of $Y|X$. So we could try to maximise the entropy of the constellation with a constraint on its power. As described in \cite{elementsOfInformation}, solving this problem using the Lagrange multipliers leads to the Maxwell--Boltzmann distribution.
\begin{align*}
p_X(x) = \frac{e^{- \lambda ||x||^2}}{\sum\limits_{x' \in \Omega} e^{- \lambda ||x'||^2}} , && \text{where } \lambda \geq 0
\end{align*}
\\
Note that if we choose $\lambda \le 0$, then we will achieve the opposite of our goal. Indeed, in this case, for the same entropy, the power is maximised if $\lambda < 0$ and minimised if $\lambda \geq 0$.
In Figure~\ref{fig:Example MB probability distribution 3d view}, examples of the Maxwell--Boltzmann distribution are shown for a 64-QAM constellation. In this figure, the length of each bar corresponds to the probability of sending a specific symbol.
\\\\
We see that, depending on our choice of $\lambda$, we will get a different probability distribution. For example, if $\lambda = 0$, we obtain a uniform distribution.  On the other hand, if $\lambda \to \infty$, we only send the points of the constellation with the smallest energy uniformly at random. \\
Notice that the Maxwell--Boltzmann distribution looks a lot like the normal distribution. Indeed, it will associate a higher probability for the symbols close to zero. The probability assigned to the other symbols will decay like a Gaussian probability density function as the symbols get away from the origin.
We also see that as $\lambda$ grows, the power of the constellation reduces. Let us compute the mutual information achieved by the Maxwell--Boltzmann distribution for $\lambda \in \{0, 0.5, 1, \dots, 10\}$. Figure~\ref{fig:MB multiple lambda not scaled} shows the mutual information over the standard deviation of the noise $\sigma$ obtained for inputs following the Maxwell--Boltzmann distribution.
\\
From Figure~\ref{fig:MB multiple lambda not scaled} we see that the parameter achieving the best mutual information is $\lambda = 0$. This corresponds exactly to the uniform distribution. It is not surprising as we have seen that as $\lambda$ grows, the energy of the constellation reduces and so does the mutual information. To obtain more relevant results, we should normalise each constellation corresponding to a fixed $\lambda$ so that it has unit energy. Note that we could also directly compare the mutual information from the Maxwell--Boltzmann and the uniform distribution by plotting the curves over SNR.\\
Figure~\ref{fig:MB multiple lambda} shows the mutual information over the standard deviation of the noise $\sigma$ obtained for inputs following the Maxwell--Boltzmann distribution. Each curve corresponds to a fix $\lambda$ and the constellation has been scaled appropriately to always have a unit energy.
\\
We immediately notice that, for a fix $\sigma$, if we properly choose the parameter $\lambda$ of the Maxwell--Boltzmann distribution, the mutual information exceeds the one given by the uniform distribution.
\\\\
To determine the best parameter for each $\sigma$, we generate the mutual information for a range of $\lambda$ as done in Figure~\ref{fig:MB multiple lambda}. Then, we search for the curve achieving the maximum mutual information for each $\sigma$. From this we are able to associate a range of standard deviation of the noise to an optimised input probability distribution.
\\
Figure \ref{fig:Comparison MB UNI QAM64} compares the curve obtained from the Maxwell--Boltzmann distribution and the curve from the uniform distribution.
\\\\
The Maxwell--Boltzmann improves largely the mutual information that we can achieve over the AWGN channel. Recall that this distribution corresponds to the distribution that maximises the entropy for a fixed energy. However, we ignored the second term of the equation of the mutual information: $I(X, Y) = H(X) - H(X | Y)$.\\
We will now explore other distributions that take into account the whole equation.

%% file: Chapters/5BA_Distribution.tex
\chapter{The Blahut--Arimoto Algorithm}

The Blahut--Arimoto algorithm \cite{ArimotoPaper}\cite{BlahutPaper} is an iterative way of computing the capacity of a fixed constellation. It takes advantage of the fact that the mutual information is a concave function in the input distribution. In the first section, we will go over the algorithm and in the second section we will study the results.

\section{Description of the algorithm}

This section was largely inspired by \cite{lectureNotesBA}. \\
Let $X$ be a discrete random variable modelling the input of the AWGN channel. The random variable $X$ takes values in $\Omega \subset V$, where $V$ is a vector space of $m$ dimensions. Let $Y$ be a discrete random variable modelling the quantized output of the channel. The random variable $Y$ takes values in $\mathcal{O} \subset V$. The input $X$ is distributed according to $p_X: \Omega \to [0,1]$. The output $Y$ is distributed according to $p_Y: \mathcal{O} \to [0,1]$.
\\\\
We rewrite the mutual information as:
\[
    I(X, Y) = \sum \limits_{y \in \mathcal{O}} \sum \limits_{x \in \Omega} p_X(x)p_{Y|X}(y|x)\log_2\frac{p_{X|Y}(x | y)}{p_X(x)} .
\]
We now want to generalise this formula to be able to iteratively maximise it.
Let $Q \in \real^{|\Omega|}$ be any set of probability distributions. Let $G_{x, y}$ be any set of conditional probability distributions.\\
We will write $Q_x$ to describe the probability associated to the symbol $x \in \Omega$. 
\\\\
We can now write the equation that estimates the mutual information as:
\[
    I^*(Q, G) = \sum \limits_{y \in \mathcal{O}} \sum \limits_{x \in \Omega} Q_x p_{Y|X}(y|x)\log_2\frac{G_{x, y}}{Q_x} .
\]
\\
In \cite{lectureNotesBA}, it is proven that $I^*(Q, G) \leq I^*(Q, p_{X|Y}(x|y))$ $\forall G$ for a fixed $Q$. Moreover, equality is achieved if and only if
\[
G_{x,y} = \frac{Q_x p_{Y|X}(y|x)}{\sum \limits_{x' \in \Omega} Q_{x'} p_{Y|X}(y|x')} = p_{X|Y}(x|y) .
\]
The proof is done by evaluating $I^*(Q, G) - I^*(Q, p_{X|Y}(x|y))$ and showing that it is less than or equal to 0.
\\\\
It is also proven that for a fixed Q:
\begin{align*} 
I^*(Q, G) &\leq \log \sum \limits_x r_x \\
r_x &= \exp(\sum \limits_y p_{Y|X}(y|x) \log G_{x, y}) ,
\end{align*}
with equality if and only if $Q_x = r_x / \sum_{x' \in \Omega} r_{x'}$. This proof is done by using the inequality $\log u \leq u - 1$ $\forall u \in \real_{>0}$.
\\\\
We can now express the capacity in terms of $Q$ and $G$:
\[
C = \max\limits_Q\max\limits_G I^*(Q, G)
\]
The idea of the Blahut--Arimoto algorithm is to first find the maximum of $I^*$ for a fixed $Q$. This gives us a new conditional probability distribution $G_{\text{new}}$. Then, we find the maximum of $I^*$ with $G_{\text{new}}$ fixed. This gives us a new probability distribution $Q_{\text{new}}$. We do these operations repeatedly until reaching convergence. Note that as the mutual information is concave in the input distribution, it can be shown that the algorithm will converge.
\\\\
To summarise, the Blahut--Arimoto algorithm consists of the following steps:
\begin{enumerate}
  \item We are given a fixed standard deviation $\sigma$ of the noise of the AWGN channel.
  \item Initialise $Q$ to an arbitrary probability distribution, with strictly positive probabilities.
  \item Compute $G_{x,y} = \frac{Q_x p_{Y|X}(y|x)}{\sum \limits_{x' \in \Omega} Q_{x'} p_{Y|X}(y|x')}$
  \item Compute $r_x = \exp(\sum \limits_y p_{Y|X}(y|x) \log G_{x, y})$ using the $G$ computed in step 3.
  \item Update $Q_x = \frac{r_x}{\sum\limits_{x' \in \Omega} r_{x'}}$.
  \item Calculate $I^*(Q, G) = \log \sum \limits_x r_x$ and compare it to the $I^*$ obtained in the last iteration. If the difference is lower than some threshold $\epsilon$, go to step 7. Otherwise, go to step 3.
  \item The input probability distribution achieving the capacity is given by $Q$.
\end{enumerate}
In the next section, we will study another method for computing the optimal distribution.

\section{Using Numerical Methods to Compute the Optimal Distribution}

\begin{figure}[ht]
    \centering
    \includegraphics[width=12cm]{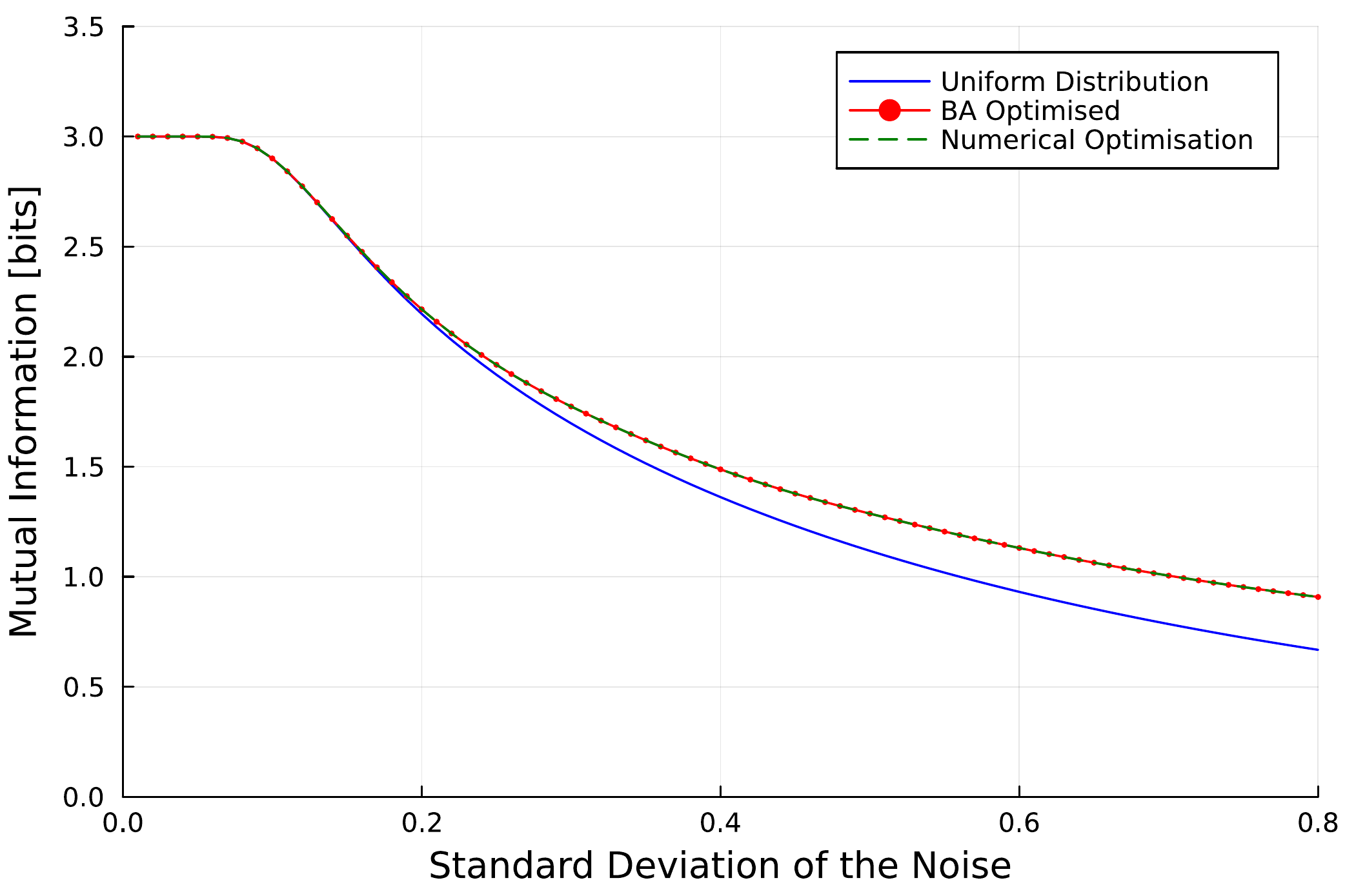}
    \caption{Comparison between the uniform distribution, the BA distribution and the numerically maximised distribution over $\sigma$ for an 8-PAM constellation}
    \label{fig:BA vs Convex Optimisation 8PAM}
\end{figure}

In the previous section, an algorithm has been described to compute the optimal distribution for a fixed constellation and a fixed standard deviation of the noise. Instead of using an iterative algorithm to maximise the mutual information, we can also use numerical methods for solving this problem. By doing so, we can also check if the Blahut--Arimoto algorithm was implemented correctly.\\
We use a library that finds the maximum of a non-linear concave function numerically. The Figure~\ref{fig:BA vs Convex Optimisation 8PAM} compares the mutual information achieved by the uniform distribution, the Blahut--Arimoto algorithm and the numerically optimised distribution over an 8-PAM constellation. This constellation was chosen to be able to increase the precision of the computations. Indeed, we can use more quantization bits and a lower threshold in the Blahut--Arimoto algorithm with the 8-PAM constellation.
\\
We can see that the curve of the Blahut--Arimoto algorithm matches the one obtained from numerical methods. Indeed, the energy of the difference between the two curves is less than $5\cdot10^{-17}$. Therefore, we have verified the correctness of our implementation of the Blahut--Arimoto algorithm. We have also confirmed that the distribution given by the Blahut--Arimoto algorithm maximises the mutual information for a fixed constellation and a fixed noise energy, and is thus optimal.

\section{Results}

\begin{figure}[ht]
    \centering
    \includegraphics[width=12cm]{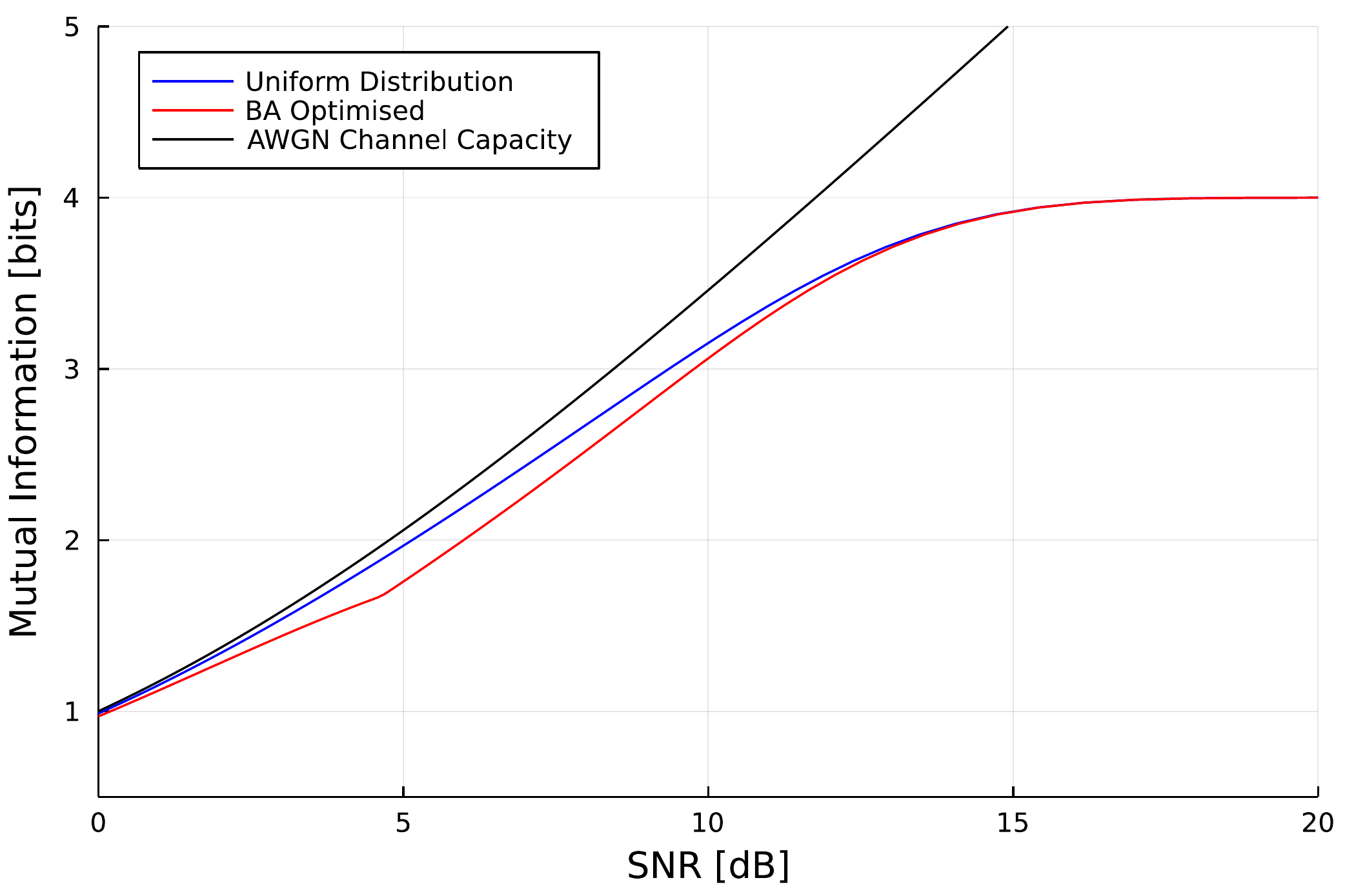}
    \caption{Comparison between the uniform distribution and the BA distribution over SNR for a 16-QAM}
    \label{fig:BA vs Uni over SNR 16QAM}
\end{figure}

\begin{figure}[ht]
    \centering
    \includegraphics[width=9cm]{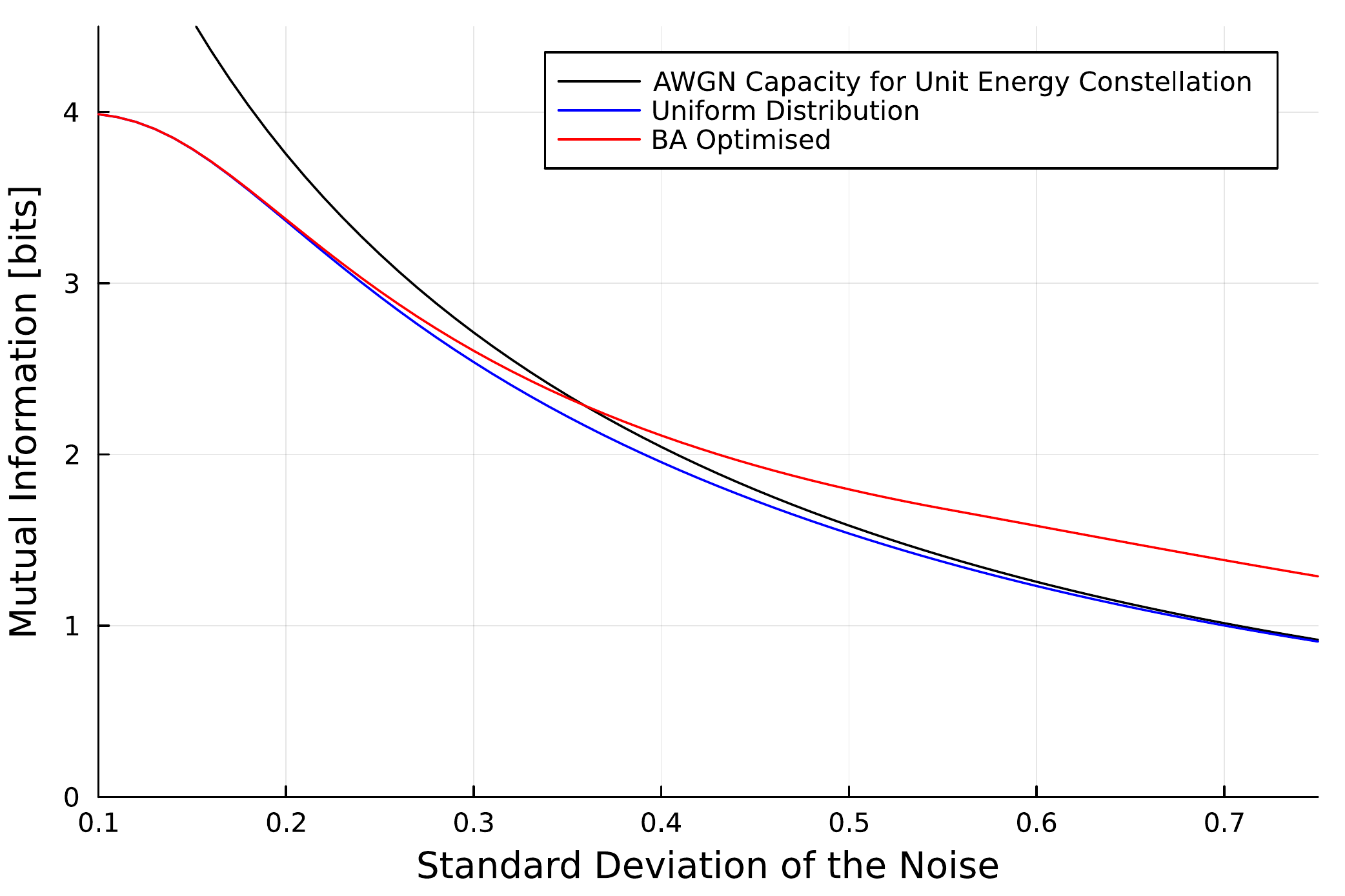}
    \caption{Comparison between the uniform and the BA distribution over $\sigma$ for a 16-QAM}
    \label{fig:BA vs Uni over sigma 16QAM}
\end{figure}

\begin{figure*}[ht]
    \centering
    \begin{subfigure}[b]{0.475\textwidth}
        \centering
        \includegraphics[width=\textwidth]{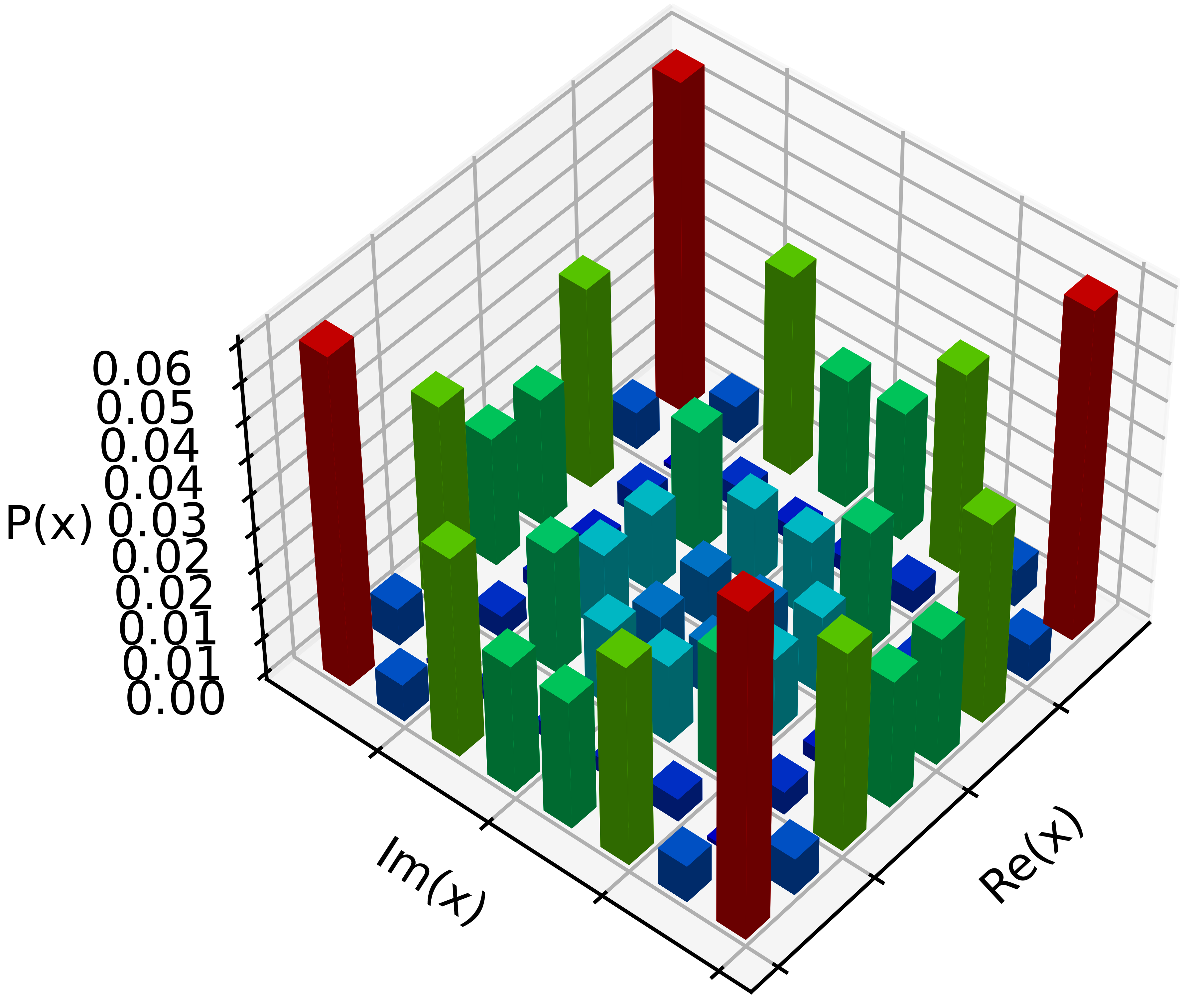}
        \caption[]%
        {{\small Distribution given by BA with $\sigma = 0.2$}}    
        \label{fig:BA Distribution Bar 0.2 sigma}
    \end{subfigure}
    \hfill
    \begin{subfigure}[b]{0.475\textwidth}  
        \centering 
        \includegraphics[width=\textwidth]{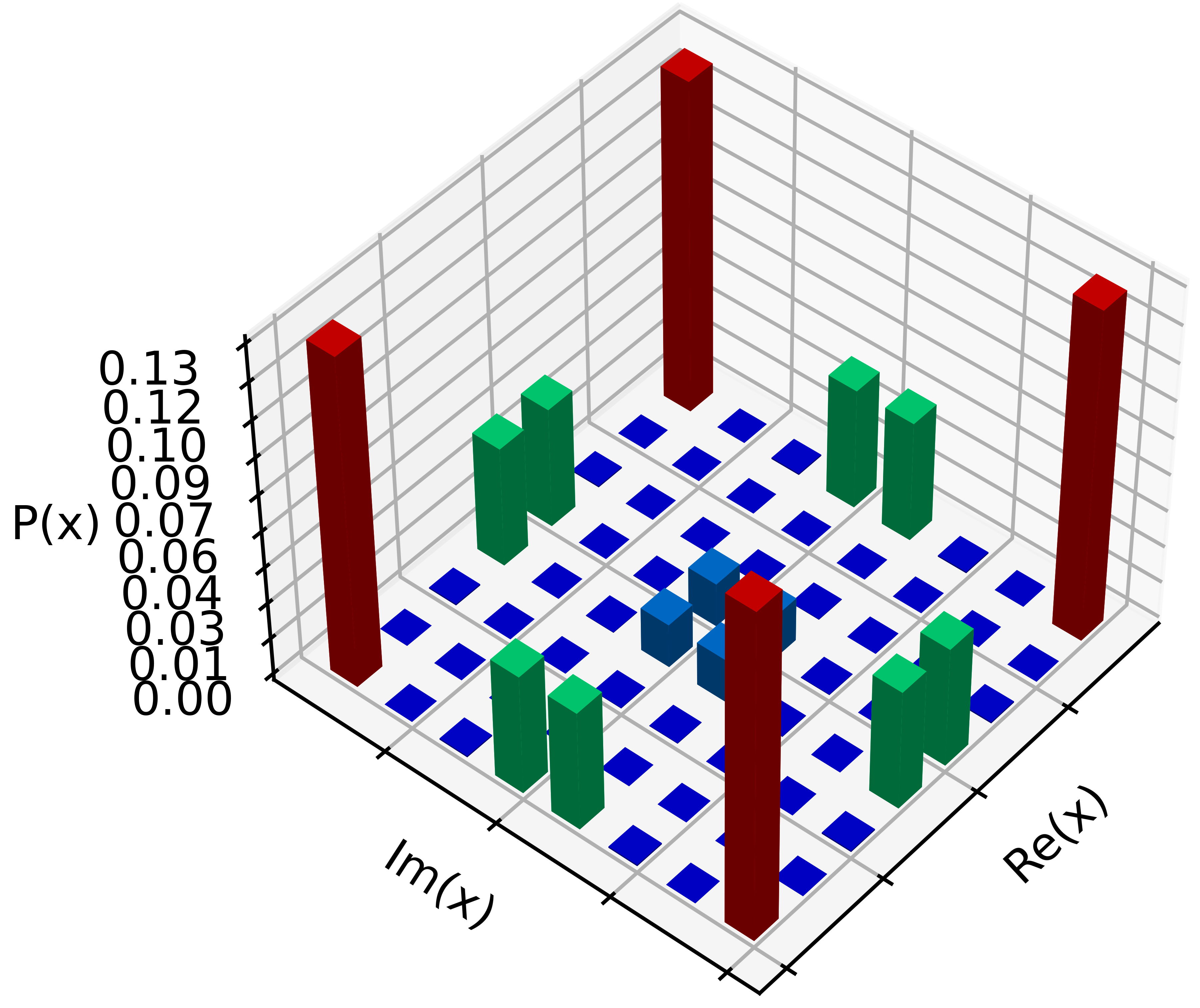}
        \caption[]%
        {{\small Distribution given by BA with $\sigma = 0.4$}}    
        \label{fig:BA Distribution Bar 0.4 sigma}
    \end{subfigure}
    \caption{Output of the Blahut--Arimoto algorithm for a 64-QAM}
    \label{fig:BA Distribution Bar}
\end{figure*}

\begin{figure}[ht]
    \centering
    \includegraphics[width=9cm]{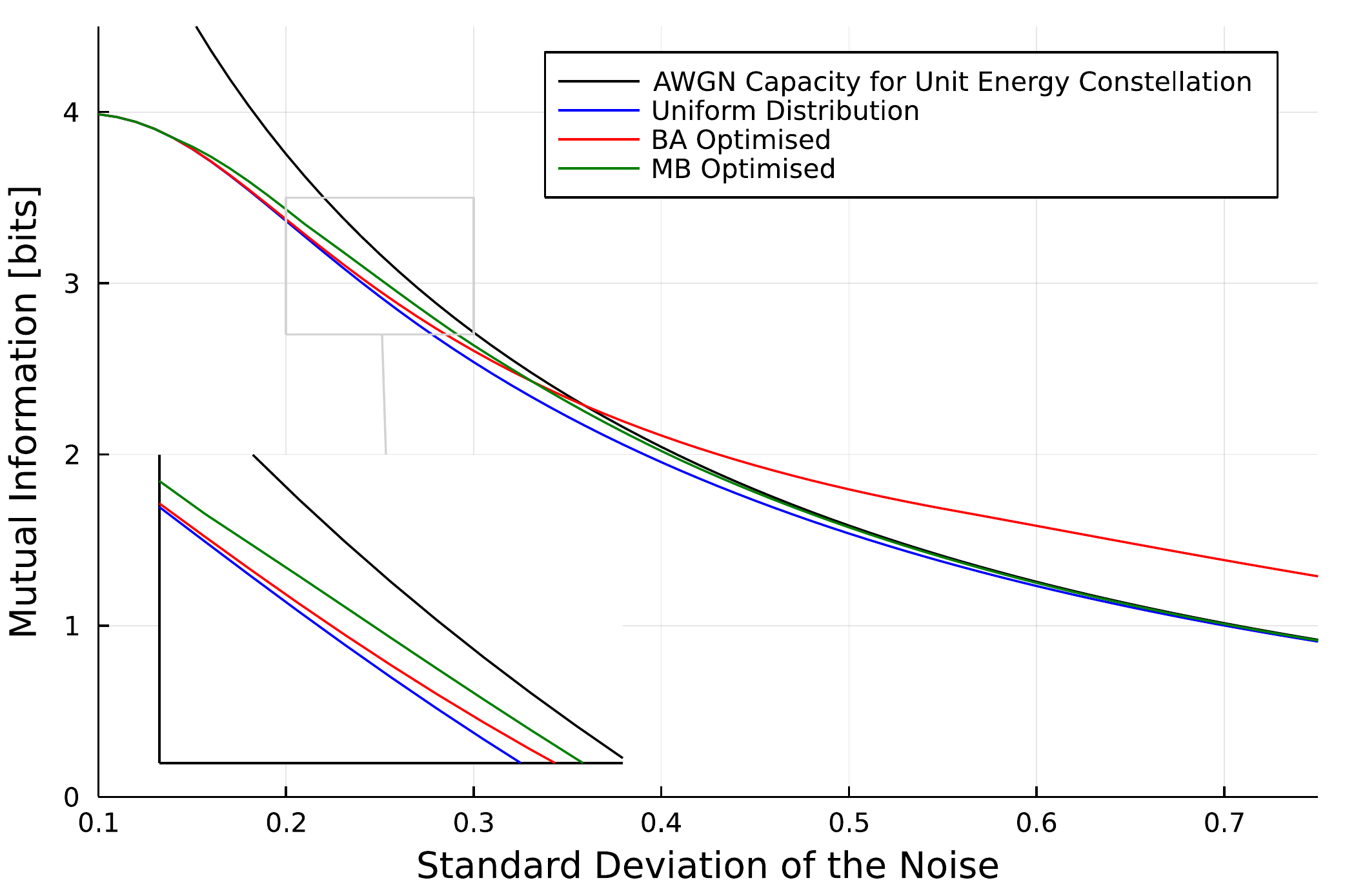}
    \caption{Comparison between the uniform, the BA and the MB distribution over $\sigma$ for a 16-QAM}
    \label{fig:BA vs Uni vs MB over sigma 16QAM}
\end{figure}

The curves of the mutual information over the SNR were not those expected. We are trying to make an improvement on the Maxwell--Boltzmann distribution by using an algorithm that gives us the \textit{optimal} distribution for a fixed constellation. Therefore, one might initially expect the curve of the mutual information given by the Blahut--Arimoto algorithm to be located between the curve of AWGN channel capacity and the curve given by the Maxwell--Boltzmann distribution.\\
\\
Figure \ref{fig:BA vs Uni over SNR 16QAM} shows a comparison between the mutual information given by the uniform distribution and the Blahut--Arimoto algorithm. The constellation used is the 16-QAM normalised to have unit energy when the input distribution is uniform.
\\
The curve obtained is counter intuitive as the mutual information of the Blahut--Arimoto algorithm is lower than the one of the uniform distribution.\\
We can also look at the mutual information obtained over the standard deviation of the noise. It might give us some hints on why we observe such a curve. \\
Figure \ref{fig:BA vs Uni over sigma 16QAM} compares the mutual information given by the uniform distribution and the Blahut--Arimoto algorithm over the standard deviation of the noise $\sigma$.
\\
First, we notice that the Blahut--Arimoto curve exceeds the AWGN capacity for unit energy constellations. Indeed, the Blahut--Arimoto algorithm computes probability distributions that result in constellations of energy bigger than one. Figure~\ref{fig:BA Distribution Bar} shows the probability distribution obtained from the Blahut--Arimoto for a 64-QAM.
\\
As the power of the noise increases, the outer points of the constellation are selected more often. It results in a constellation having a higher energy.\\
In Figure~\ref{fig:BA vs Uni over SNR 16QAM}, Blahut--Arimoto is compared to the uniform distribution over SNR. Thus, the difference of energy between the two constellation is taken into account when comparing the two curves. On the contrary, when we do the same plot over $\sigma$, the energy of the constellations are not normalised. This is why we observe a discontinuity in the curve given by Blahut--Arimoto over SNR in \ref{fig:BA vs Uni over SNR 16QAM} and not in the same curve in \ref{fig:BA vs Uni over sigma 16QAM} over $\sigma$.\\\\
By looking at Figure~\ref{fig:BA vs Uni over sigma 16QAM}, we can verify that Blahut--Arimoto achieves a better mutual information than the one given by the uniform distribution for a \textit{fixed} constellation and a fixed $\sigma$. One might be tempted to compare the mutual information given by the Maxwell--Boltzmann distribution that we calculated earlier. Figure~\ref{fig:BA vs Uni vs MB over sigma 16QAM} shows a comparison between the mutual information given by the Blahut--Arimoto, the uniform and the Maxwell--Boltzmann distribution.
\\
We notice that in Figure~\ref{fig:BA vs Uni vs MB over sigma 16QAM}, for approximately $\sigma \in [0.16, 0.32]$, the mutual information given by the Maxwell--Boltzmann distribution exceeds the one given by the Blahut--Arimoto algorithm. We can explain this observation by arguing that the constellation given to the Blahut--Arimoto algorithm is \textit{fixed}. In our case, we use a 16-QAM constellation that is normalised so that, when the input distribution is uniform, the constellation has unit energy. Recall how each point of the Maxwell--Boltzmann distribution is computed in Section~4. We apply a gain to the input alphabet so that, for each $\sigma$, the constellation given by Maxwell--Boltzmann has unit energy. So the Blahut--Arimoto algorithm is not optimising the input distribution for the same constellation as the one used with Maxwell--Boltzmann. Therefore, the graph in Figure~\ref{fig:BA vs Uni vs MB over sigma 16QAM} can not help us compare the mutual information achieved by Maxwell--Boltzmann distribution and the one achieved by the Blahut--Arimoto algorithm. \\\\
Nevertheless, recall that the Maxwell--Boltzmann distribution minimises the energy for a fixed entropy if the parameter of the distribution is non-negative. We also know that the Maxwell--Boltzmann distribution gives a probability distribution that has the opposite shape of a Gaussian distribution if $\lambda < 0$. Here we observe that the distribution given by the Blahut--Arimoto algorithm associates a bigger probability to the outer points in general (see Figure~\ref{fig:BA Distribution Bar}). So out of curiosity, we can compare the mutual information given by the Maxwell--Boltzmann distribution with non-positive parameters to the one given by the Blahut--Arimoto algorithm.

\section{Comparison Between High Energy Maxwell--Boltzmann Distributions and Blahut--Arimoto Distributions}

\begin{figure}[ht]
    \centering
    \includegraphics[width=9cm]{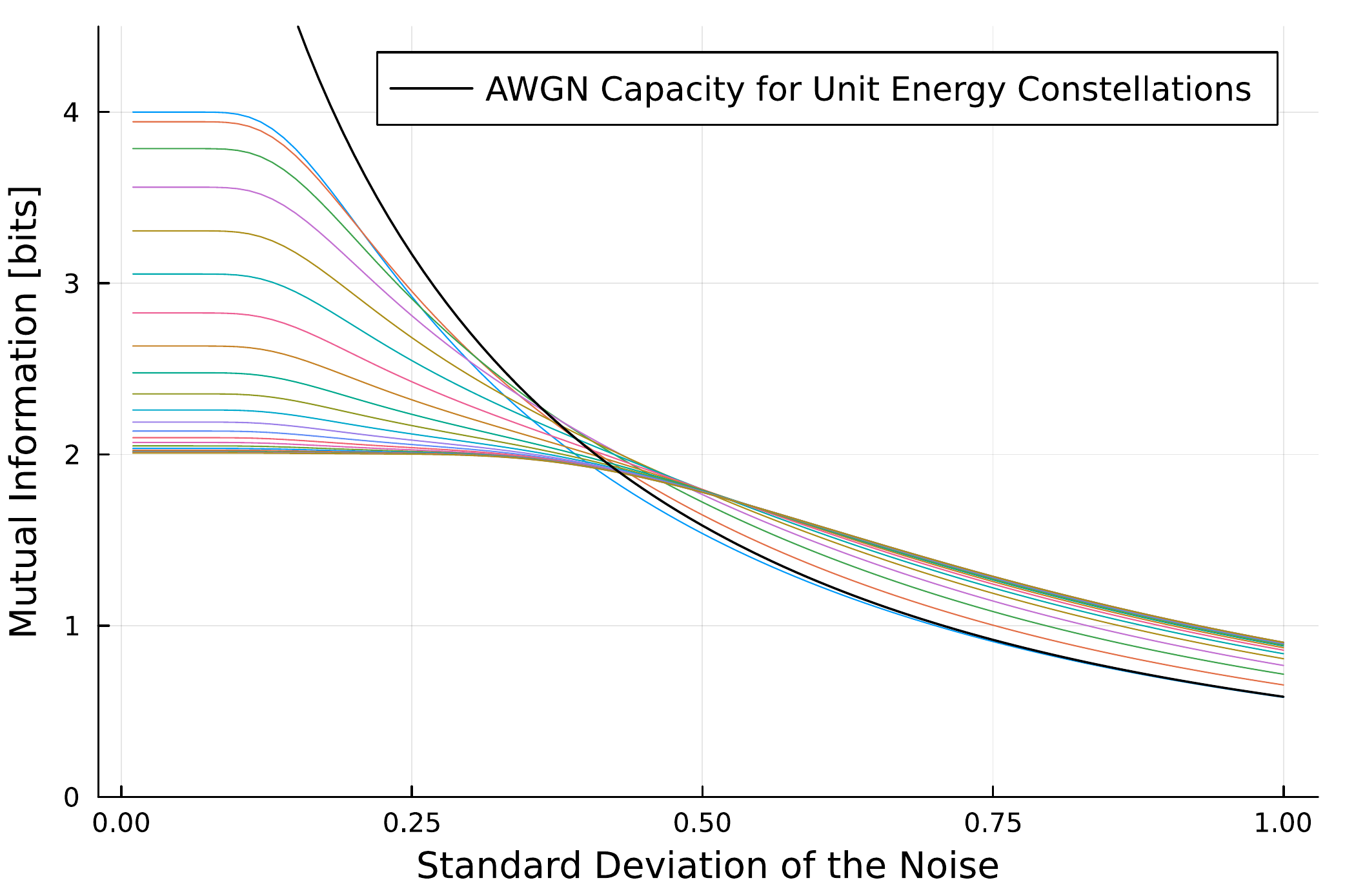}
    \caption{MB distribution for $\lambda \in \{0, -0.5, \dots, -10\}$ over $\sigma$ for a 16-QAM}
    \label{fig:MB multiple lambda max energy}
\end{figure}

\begin{figure}[ht]
    \centering
    \includegraphics[width=10cm]{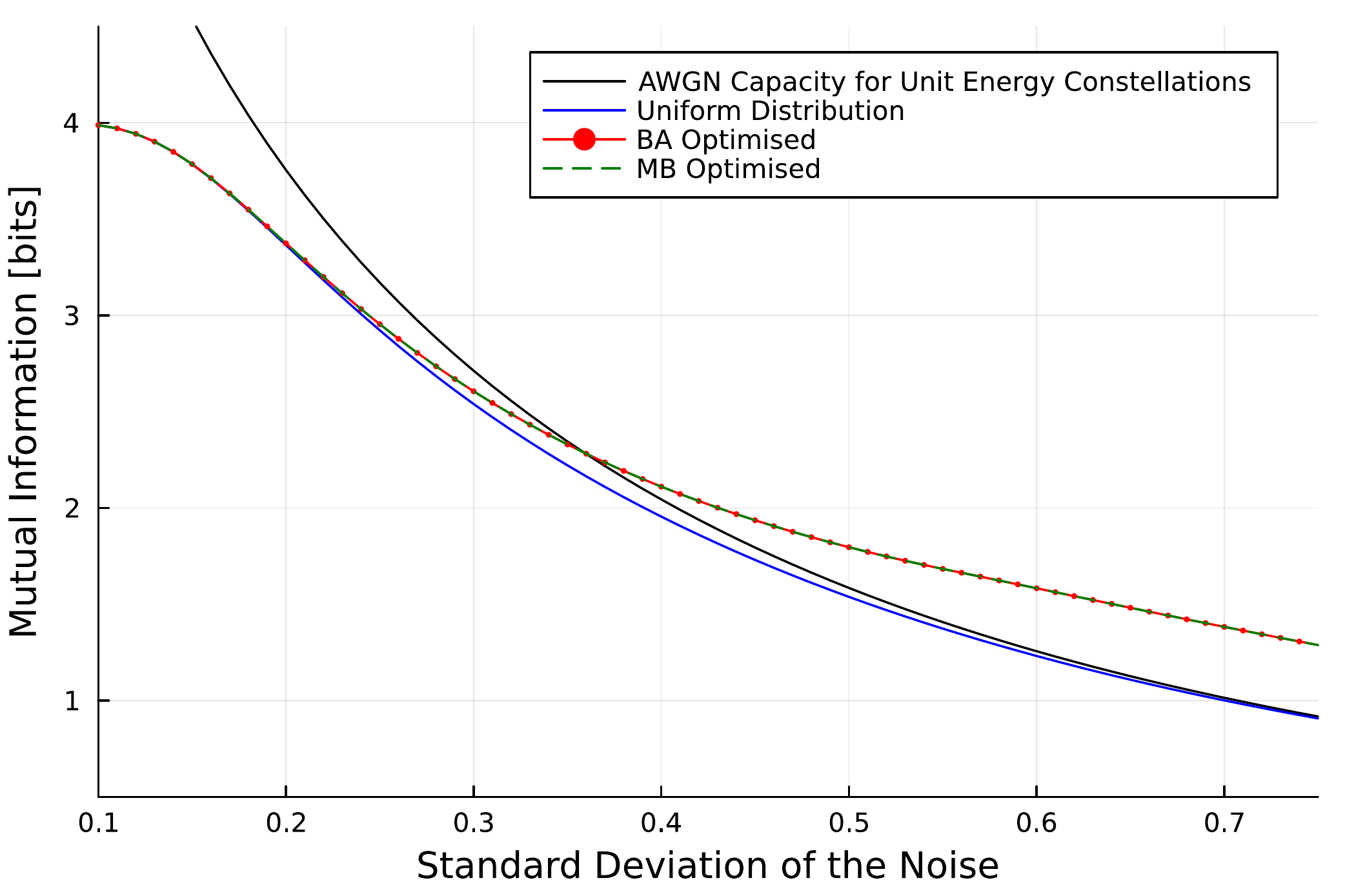}
    \caption{MI of a 16-QAM constellation over $\sigma$ given by the MB distribution with $\lambda \leq 0$ and the Blahut--Arimoto algorithm}
    \label{fig:MB energy max vs BA}
\end{figure}

\begin{figure}[ht]
    \centering
    \includegraphics[width=10cm]{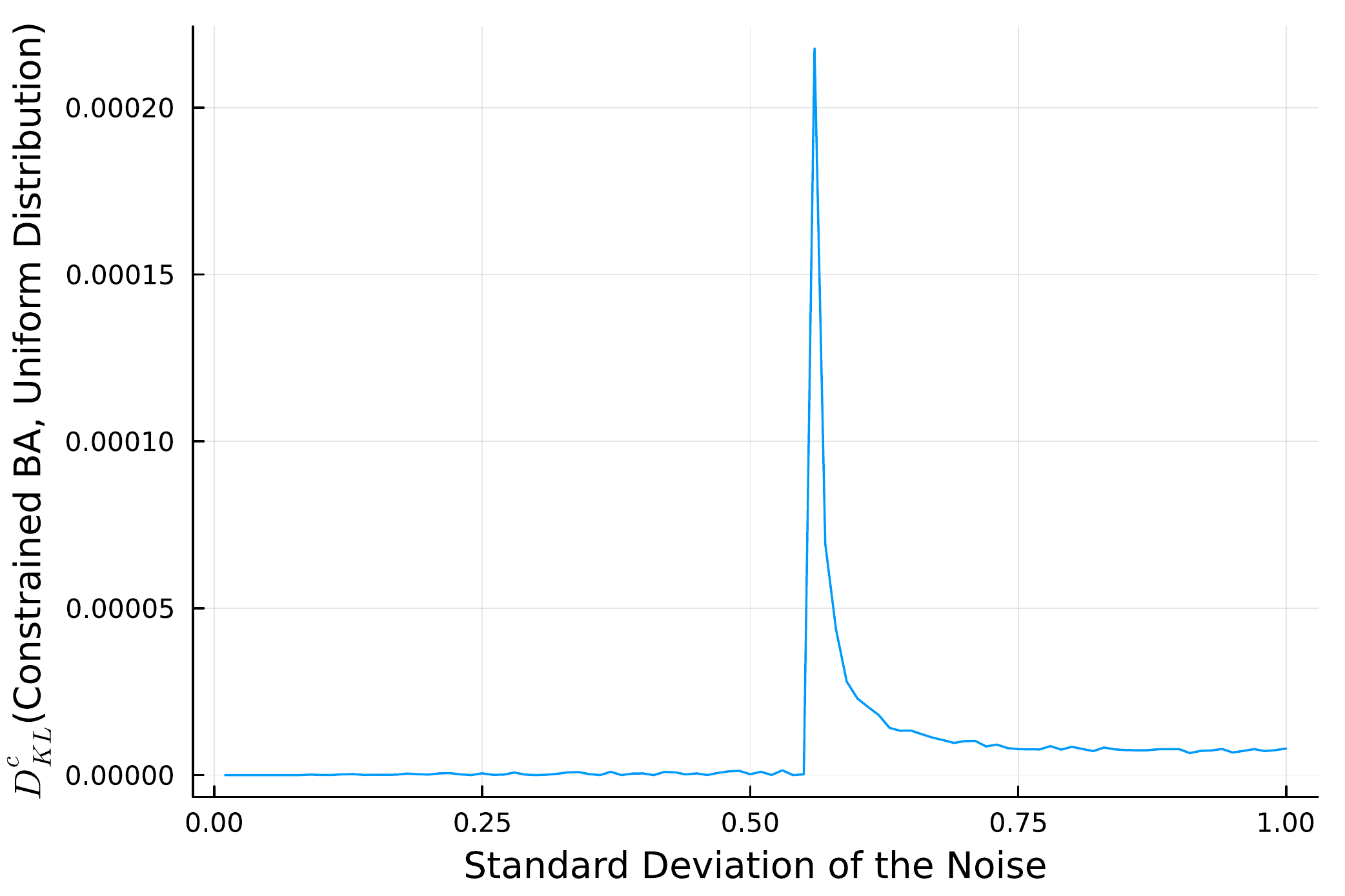}
    \caption{KL divergence between the Maxwell--Boltzmann and the Blahut--Arimoto distribution over $\sigma$}
    \label{fig:KL MB energy max vs BA}
\end{figure}

Figure~\ref{fig:MB multiple lambda max energy} shows the mutual information achieved by the Maxwell--Boltzmann distribution for $\lambda \in \{0, -0.5, \dots, -10\}$. A 16-QAM constellation is used. It has been normalised to have unit energy if the input probability distribution is uniform. No gain is applied to the system.
\\
We see that the curves crossover themselves and that for a fixed $\sigma$ there is one choice for $\lambda$ that achieves the best mutual information. Notice that, to get a similar curve, we have to apply a gain to the input when $\lambda$ is positive.\\
Correspondingly to the curve of the mutual information given by Blahut--Arimoto, the curve obtained by taking the best $\lambda < 0$ for each $\sigma$ crosses the AWGN channel capacity of unit energy constellations. Figure~\ref{fig:MB energy max vs BA} compares the mutual information obtained from Maxwell--Boltzmann with $\lambda \leq 0$ to the one obtained from Blahut--Arimoto. The threshold of the Blahut--Arimoto is set to $10^{-7}$. Concerning the Maxwell--Boltzmann distribution, we choose $\lambda = -\exp(v_i)+1$, where $v_i$ is taken uniformly in the interval $[0, 4.5]$ and $i = 1,\dots, 1500$. A fixed 16-QAM constellation is used. It has unit energy when the input distribution is uniform.\\
\\
We see that the curve obtained from the Maxwell--Botzmann distribution is very close to the one obtained from the Blahut--Arimoto algorithm. Indeed, the difference of energy of the two curves is less than $6.2 \cdot 10^{-12}$. So, up to numerical precision, the mutual information obtained from the two distributions is the same.\\
Figure \ref{fig:KL MB energy max vs BA}, shows the commutative KL divergence of the distribution obtained from Blahut--Arimoto and the one given by Maxwell--Boltzmann.
\\
Overall, the two distributions are very similar. Nevertheless, at $\sigma = 0.56$, we observe a spike in the KL divergence. This discontinuity is due to the fact that for $\sigma = 0.55$, the best distribution from Maxwell--Boltzmann is given by $\lambda \approx -5.8$ and for $\sigma = 0.56$, it is given by $\lambda \approx -46.2$. This is a relatively large difference of $\lambda$. Note, that even though there is a spike in the KL divergence, the difference between the two distributions is still very low.
\\\\
We were able to implement an algorithm to get the optimal distribution for a fixed constellation. However, we have seen that there is no constraint on the energy that this constellation can have. In most practical applications, we want to maximise the mutual information over the SNR, not over the energy of the noise. Therefore, in the next section, we will study a constrained version of the Blahut--Arimoto algorithm that enables us to have the control on the energy of the optimised constellation.

%% file: Chapters/6BA_Constrained.tex
\chapter{The Constrained Blahut--Arimoto algorithm}

\begin{figure}[ht]
    \centering
    \includegraphics[width=9cm]{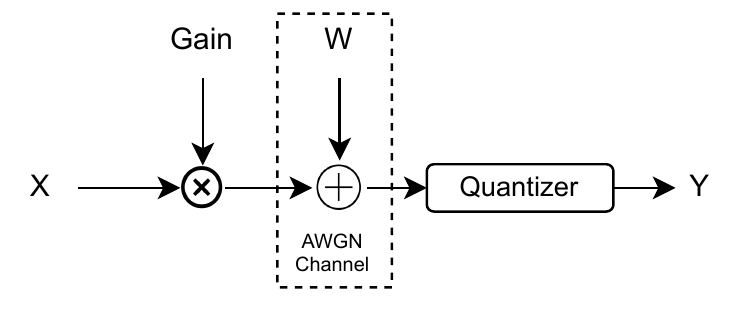}
    \caption{The quantized AWGN channel with a gain applied on the input}
    \label{fig:diagram AWGN with gain}
\end{figure}

In \cite{constrainedBA} an algorithm is described that adapts the Blahut--Arimoto algorithm to add a constraint on the power of the constellation. The idea is to add a parameter $\alpha \in \real_{>0}$ that represents a gain applied to the input signal before the AWGN channel. A diagram in Figure~\ref{fig:diagram AWGN with gain} shows the new system.
\\
The problem we want to solve is to maximise the mutual information under some constraints. The most important one is that the power of the constellation is equal to a constant $\mathcal{P}$. 

\section{Description of the Algorithm}

We describe the algorithm into two parts. One part maximises the mutual information for a fixed $\alpha$ with the constraint that the power of the constellation must equal $\mathcal{P}$. The other part, selects different gains and uses part one to find the best mutual information for each gain. It then returns the gain associated to the input distribution that gives the maximum mutual information over all.
\\\\
We summarise this process as follows:
\begin{enumerate}
  \item Choose some $\alpha \in \real_{>0}$.
  \item Compute the maximal mutual information achievable with $\alpha$ fixed. This process also gives us an optimal input probability distribution for this $\alpha$.
  \item Choose the next $\alpha$ and go to step 2. Stop and return the optimal probability distribution if we determine that a satisfying precision has been reached.
\end{enumerate}

We will now focus on the second step of the algorithm.

\subsection{The modified Blahut--Arimoto Algorithm}

\begin{figure}[ht]
    \centering
    \includegraphics[width=12cm]{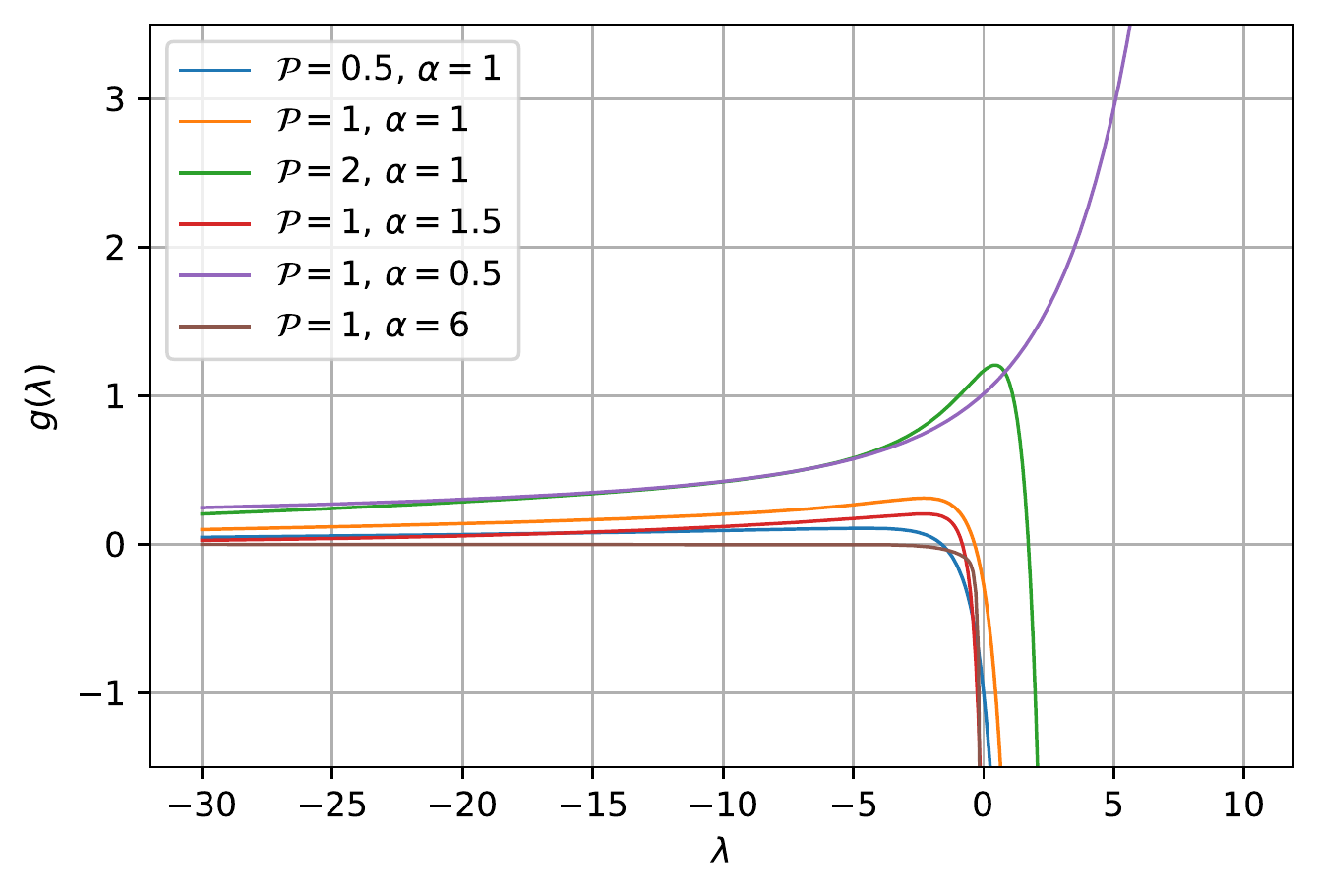}
    \caption{g as a function of $\lambda$ for an 8-PAM constellation}
    \label{fig:g(lambda) for 8-PAM}
\end{figure}

Let $X$ be a discrete random variable modelling the input of the AWGN channel. The random variable $X$ takes values in $\Omega \subset V$, where $V$ is a vector space of $m$ dimensions. Let $Y$ be a discrete random variable modelling the quantized output of the channel. The random variable $Y$ takes values in $\mathcal{O} \subset V$. The input $X$ is distributed according to $p_X: \Omega \to [0,1]$. \\
We want to maximise the mutual information $I(X, Y)$ with the following constraint:
\[
    \sum\limits_{x \in \Omega} p_X(x) \alpha^2 ||x||^2 = \mathcal{P} 
\]
In \cite{constrainedBA}, a modified version of the Blahut--Arimoto algorithm is defined. The goal of the algorithm is to maximise the mutual information by changing the input distribution. The difference with the original Blahut--Arimoto algorithm is that we now have to account for the gain added to the system. The modified Blahut--Arimoto consists of the following steps:
\begin{enumerate}
  \item Fix $\alpha$ and choose an arbitrary probability distribution $p_X$.
  \item Evaluate $T_x = \mathbb{E}_Y\left[\frac{p(X = x | Y) \log_2(p(X = x | Y))}{p_X(x)}\right]$.
  \item Compute $p_X = \argmax_{r}\sum\limits_{x \in \Omega} r_x [\log_2(\frac{1}{r_x}) + T_x]$ under the constraints $\sum_x r_x = 1$, $r_x > 0$ $\forall x$ and $\sum_x p_X(x)\alpha^2||x||^2 = \mathcal{P}$.
  \item If we have not reached convergence, go to step 2. Otherwise return the last input probability distribution $p_X$ computed.
\end{enumerate}
To solve the third step, we use the method of Lagrange multipliers. As shown in \cite{constrainedBA}, we obtain the following equations:
\begin{align}
&p_X(x) = \frac{2^{T_x + \lambda \alpha^2 ||x||^2}}{\sum \limits_{x' \in \Omega}2^{T_{x'} + \lambda \alpha^2 ||x'||^2}} \label{eqn:constrained ba px}\\
&T_x = \mathbb{E}_Y\left[\frac{p(X = x | Y) \log_2(p(X = x | Y))}{p_X(x)}\right] \label{eqn:constrained ba ti}\\
&\sum\limits_{x \in \Omega}(\mathcal{P} - \alpha^2||x||^2)2^{T_x}\cdot 2^{\lambda \alpha^2||x||^2} = 0 \label{eqn:constrained ba lambda}
\end{align}
Note that in order to solve \eqref{eqn:constrained ba px}, we first have to know the value of $\lambda$. Clearly, the left hand side of \eqref{eqn:constrained ba lambda} goes to 0 as $\lambda$ goes to minus infinity. But this is not the solution we are looking for. Let us take a look at the plot of the function $g(\lambda) = \sum\limits_{x \in \Omega}(\mathcal{P} - \alpha^2||x||^2)2^{T_x}\cdot 2^{\lambda \alpha^2||x||^2}$ in Figure~\ref{fig:g(lambda) for 8-PAM}. 
\\
In Figure~\ref{fig:g(lambda) for 8-PAM}, we see that $g$ does not always cross the horizontal axis depending on the parameters of the curve. Intuitively, if the ratio $\mathcal{P}$ over $\alpha$ is too big, the function $g(\lambda)$ is strictly bigger than zero. Conversely, if the ratio is too low, $g(\lambda) < 0$ $\forall \lambda$. Knowing this, we choose to fix $\mathcal{P} = 1$ and we choose, for the moment, to restrain the gain $\alpha$ to the interval $[0.5, 5]$.
\\\\
We have found an interval that contains the gains for which $g$ crosses the horizontal axis. We can now use numerical methods to find the zeros of $g$. From this, we can compute $\lambda$ that satisfies \eqref{eqn:constrained ba lambda}. Finally, knowing the value of $\lambda$, we can directly compute $p_X$.
\\\\
To summarise, to find the maximum mutual information achievable for a given $\alpha$ and $\mathcal{P}$, we apply the steps one to four as shown above. We solve the third step by using the method of Lagrange multipliers. We don't have any closed form formula for solving the equations given by the method of Lagrange multipliers, so we use numerical methods.

\subsection{Maximising the Mutual Information over the gain}

In the previous subsection, we found that given a power constraint, not all gains yield to solvable equations in the modified Blahut--Arimoto algorithm. Therefore, we choose to use, for the moment, $\alpha \in [0.5; 5]$.\\
One option would be to choose a small $\delta \in \real$ and run iteratively the modified Blahut--Arimoto algorithm with $\alpha = \alpha_{\text{min}} + k\delta$, where $k \in S = \{ 0, 1, \dots, \lfloor \frac{\alpha_{\text{max}} - \alpha_{\text{min}}}{\delta} \rfloor \}$ and $\alpha_{\text{min}} = 0.5$, $\alpha_{\text{max}} = 5$. In fact, this is not very efficient. If we want to be very precise, we have to choose $\lambda$ very small. This results in a long computation time.\\
A better option would be to repeatedly update the set of gains for which we want to maximise the mutual information. To this end, we choose a number of iterations $n$.
We then find the gain $\alpha_{\text{cap}}$ that maximises the mutual information that is in the set $\{ \alpha_{\text{min}} + k \frac{\alpha_{\text{max}} - \alpha_{\text{min}}}{n - 1} : k \in \{ 0, \dots, n-1 \} \}$. We can now do the same operations with a new set of gains in a smaller interval. Indeed, by updating $\alpha_{\text{min}}$ and $\alpha_{\text{max}}$ according to the following
\begin{align*}
\alpha_{\text{max}}' &= \alpha_{\text{cap}} + \frac{\alpha_{\text{max}} - \alpha_{\text{min}}}{n-1}\\
\alpha_{\text{min}}' &= \alpha_{\text{cap}} - \frac{\alpha_{\text{max}} - \alpha_{\text{min}}}{n-1}\\
\end{align*}
We get a smaller interval that contains the optimal gain. We call the \textit{depth} the number of times we update the set of gains. This algorithm results in a logarithmic search instead of a linear search for the optimal gain.

\section{Mutual Information}

\begin{figure}[ht]
    \centering
    \includegraphics[width=10cm]{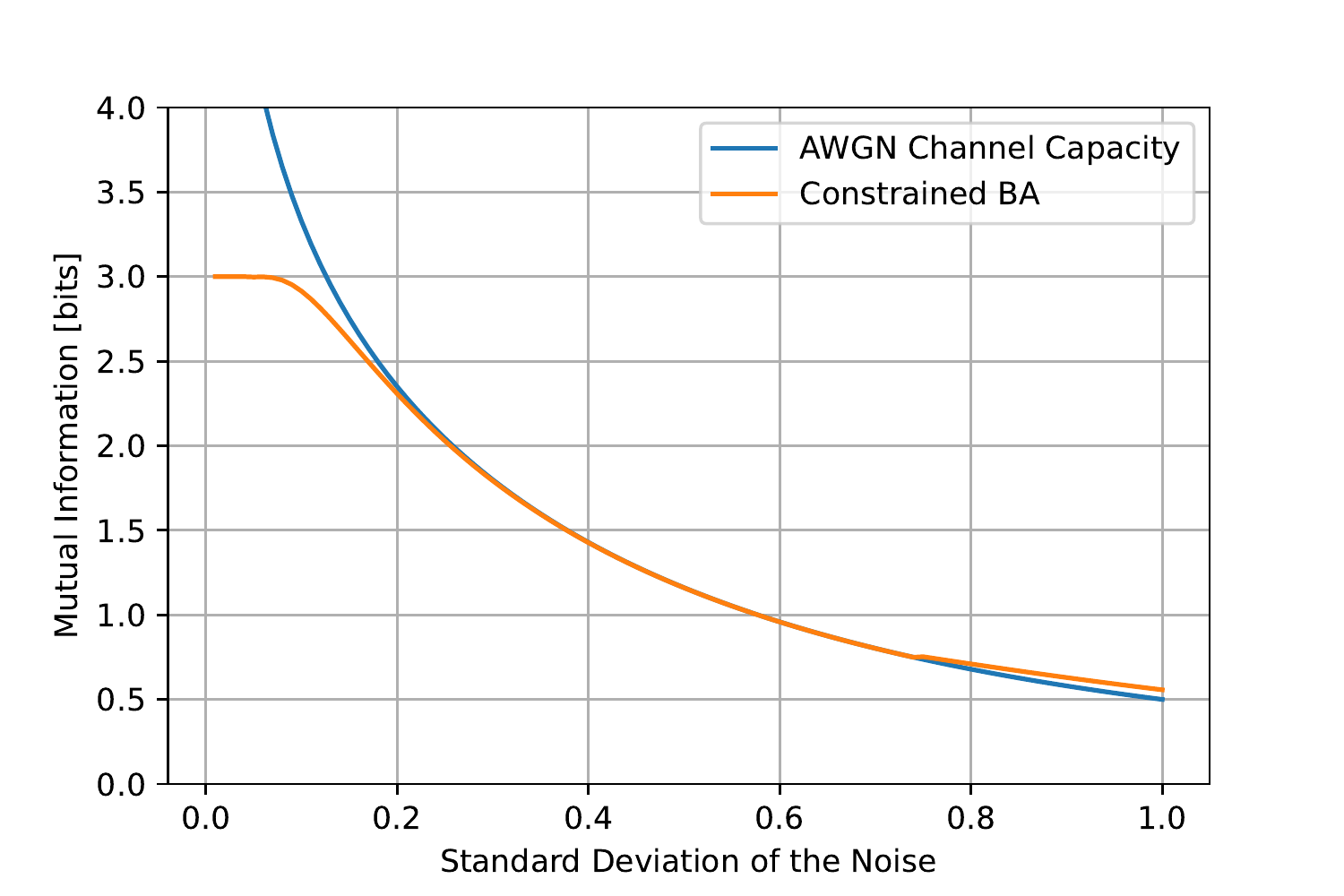}
    \caption{MI of a 8-PAM constellation given by the constrained BA with $\alpha_{\text{max}} = 5$}
    \label{fig:8-PAM Constrained BA alpha max 5}
\end{figure}

\begin{figure}[ht]
    \centering
    \includegraphics[width=10cm]{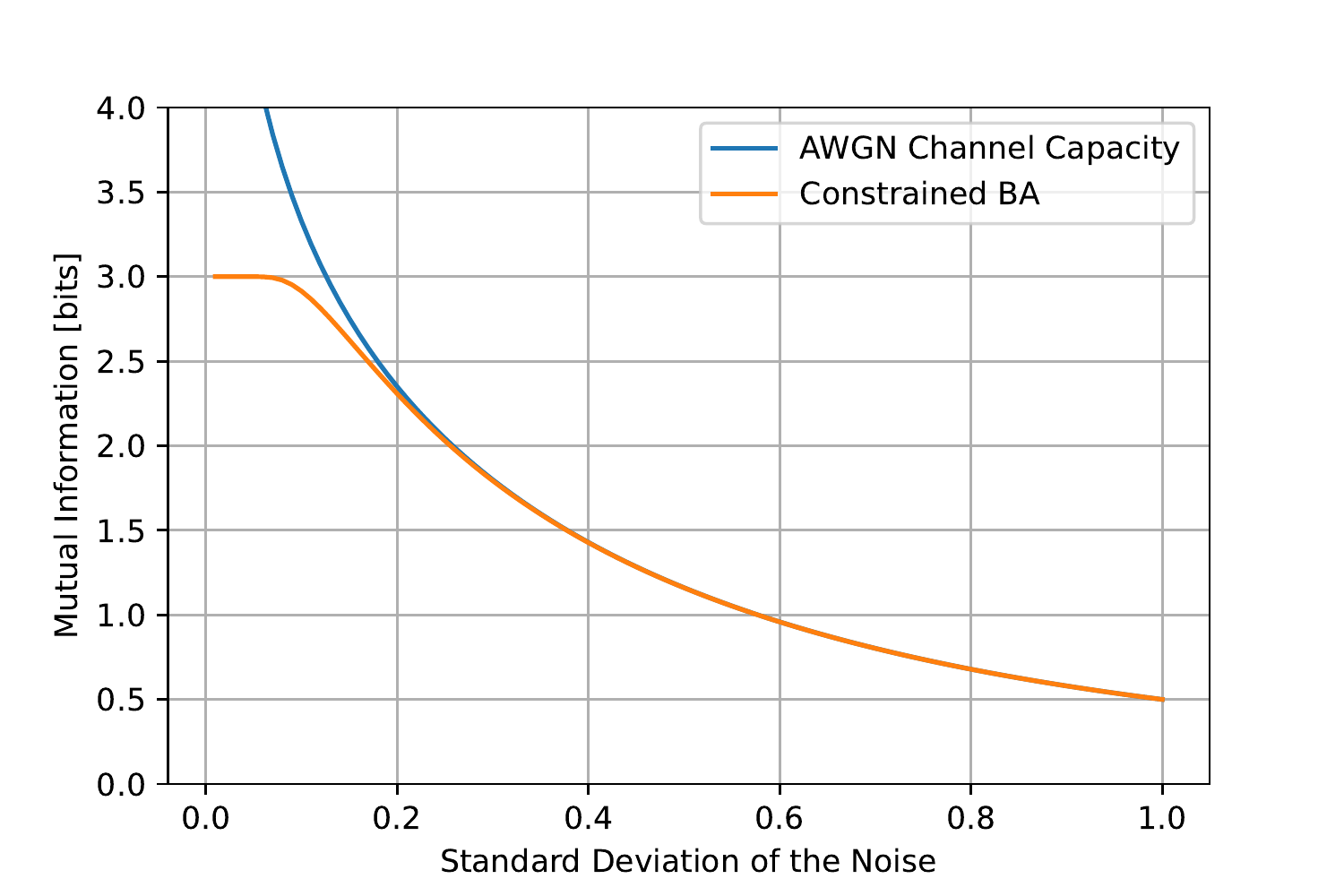}
    \caption{MI of a 8-PAM constellation given by the constrained BA with $\alpha_{\text{max}} = 4$}
    \label{fig:8-PAM Constrained BA alpha max 4}
\end{figure}

In Figure~\ref{fig:8-PAM Constrained BA alpha max 5} is shown the mutual information over the standard deviation of the noise for a 8-PAM constellation. It compares the mutual information given by the constrained Blahut--Arimoto against the AWGN channel capacity. We have used $\alpha_{\text{min}} = 0.5$ and $\alpha_{\text{max}} = 5$ for the reasons discussed above. The stopping criterion in the modified Blahut--Arimoto algorithm is that the difference of mutual information between two iterations is less or equal to 1e-6. The depth of the search for the best gain is five.
\\
We observe that the mutual information of the constrained Blahut--Arimoto goes over the upper bound. This means that the constellation does not have unit energy. At the same time, the goal of the constrained Blahut--Arimoto algorithm was to force the constellation to have unit energy. So we conclude that there must either be a bug in the program, either that there has been a problem regarding the convergence of the modified Blahut--Arimoto algorithm.\\
Most of the graphic looks right. It is only for high standard deviations that there is a problem. In fact, there seems to be a discontinuity at $\sigma = 0.75$. This might be a hint that the program is working but that for some standard deviations, the modified Blahut--Arimoto algorithm does not converge.\\\\
Recall that we made the imprecise choice of initialising $\alpha_{\text{max}} = 5$. Moreover, we can calculate that for $\alpha = 5$, the function $g$ does not cross the horizontal axis. Now, if we look at the gains outputted by the constrained Blahut--Arimoto algorithm when computing the curve in \ref{fig:8-PAM Constrained BA alpha max 5}, we can see that at exactly $\sigma = 0.75$, the gain jumps from $\alpha \approx 2.67$ to $\alpha = 5$.\\
We conclude that we have to choose a thinner range of gains when computing the optimal input distribution. We now choose gains from $\alpha_{\text{min}} = 0.5$ to $\alpha_{\text{max}}= 4$. Figure~\ref{fig:8-PAM Constrained BA alpha max 4} shows the same plot as above, but with the range of gains being more limited.
\\
As we can see in Figure~\ref{fig:8-PAM Constrained BA alpha max 4}, we are able to get the curve of the capacity of the 8-PAM constellation. Note that the optimal gains did not get close to $\alpha_{\text{max}}= 4$. \\
So we found an interval that is large enough to contain the optimal gains and thin enough to exclude gains that would lead to invalid solutions.\\\\
The curve obtained looks a lot like the curve we got from the Maxwell--Boltzmann distribution. We will now compare the mutual information obtained from the constrained Blahut--Arimoto with other input distributions.

\section{Comparison Against Other Distributions}

\begin{figure*}[ht]
    \centering
    \begin{subfigure}[b]{0.425\textwidth}
        \centering
        \includegraphics[width=\textwidth]{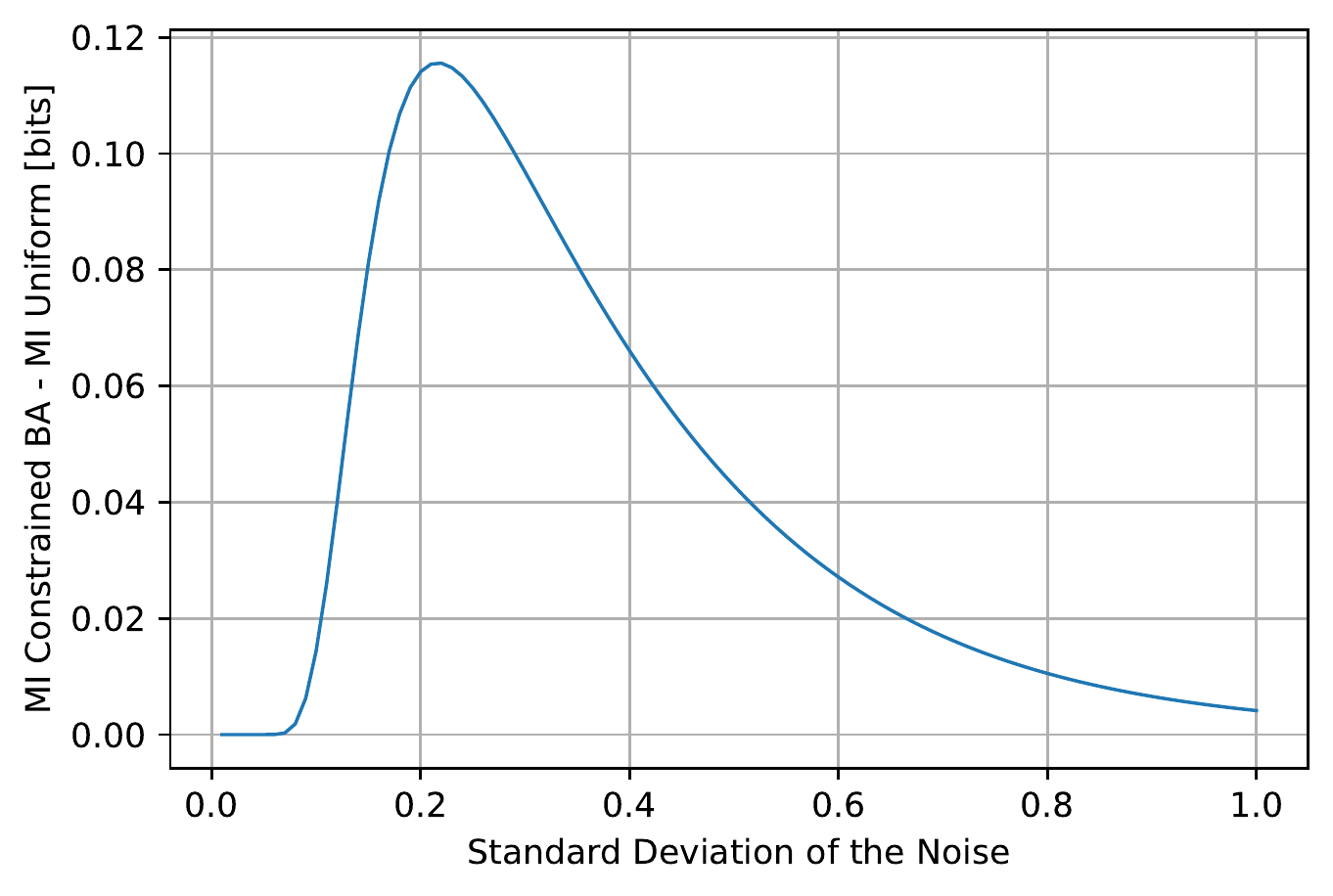}
        \caption[]%
        {{\small Difference of mutual information between the constrained BA and the uniform distribution}}    
        \label{fig:CBA diff unif}
    \end{subfigure}
    \hfill
    \begin{subfigure}[b]{0.45\textwidth}  
        \centering 
        \includegraphics[width=\textwidth]{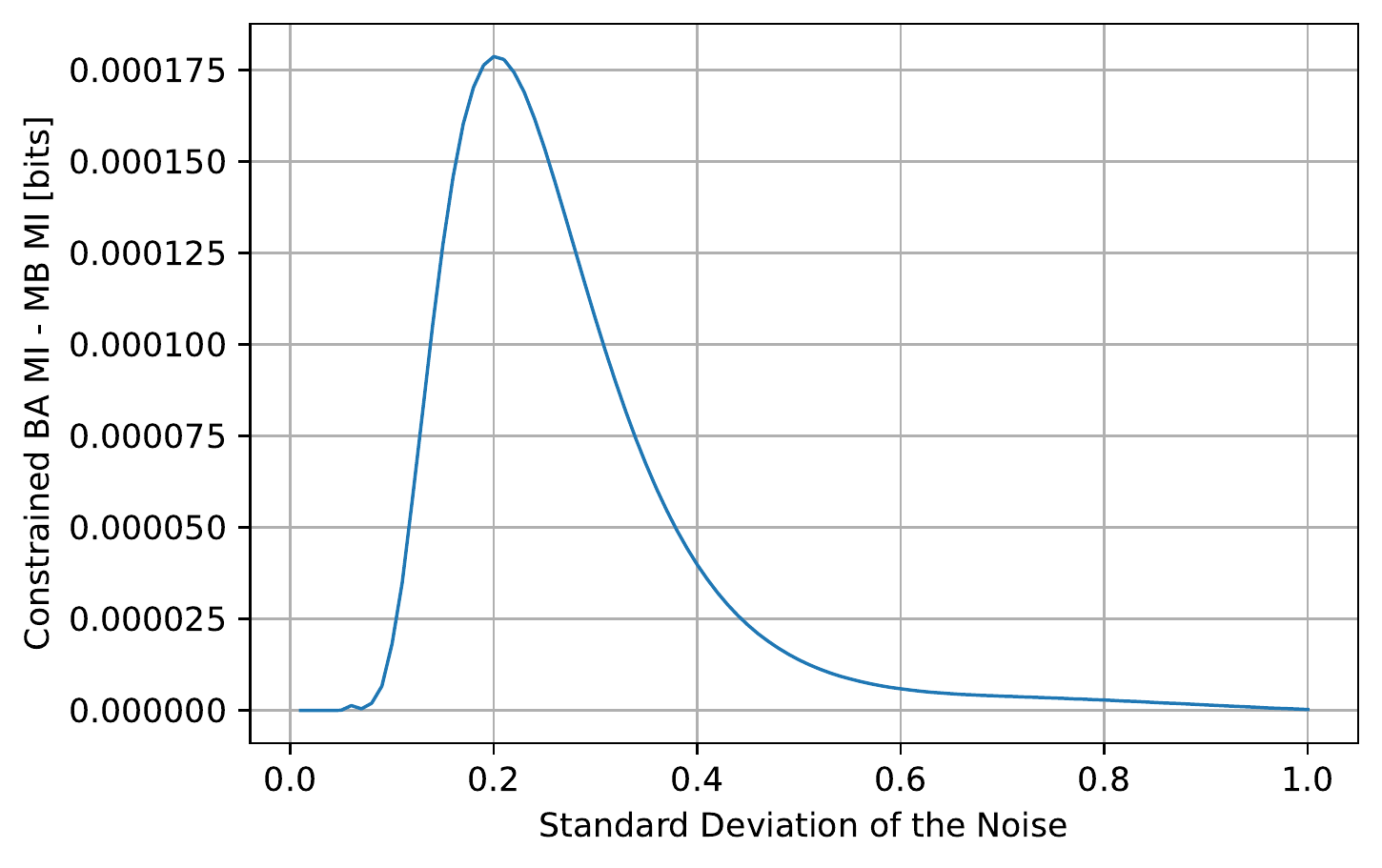}
        \caption[]%
        {{\small Difference of mutual information between the constrained BA and the MB distributions}}    
        \label{fig:CBA diff MB}
    \end{subfigure}
    \caption{Comparison of the constrained BA against the uniform and the MB distributions}
    \label{fig:CBA diff}
\end{figure*}

\begin{figure*}[ht]
    \centering
    \begin{subfigure}[b]{0.425\textwidth}
        \centering
        \includegraphics[width=\textwidth]{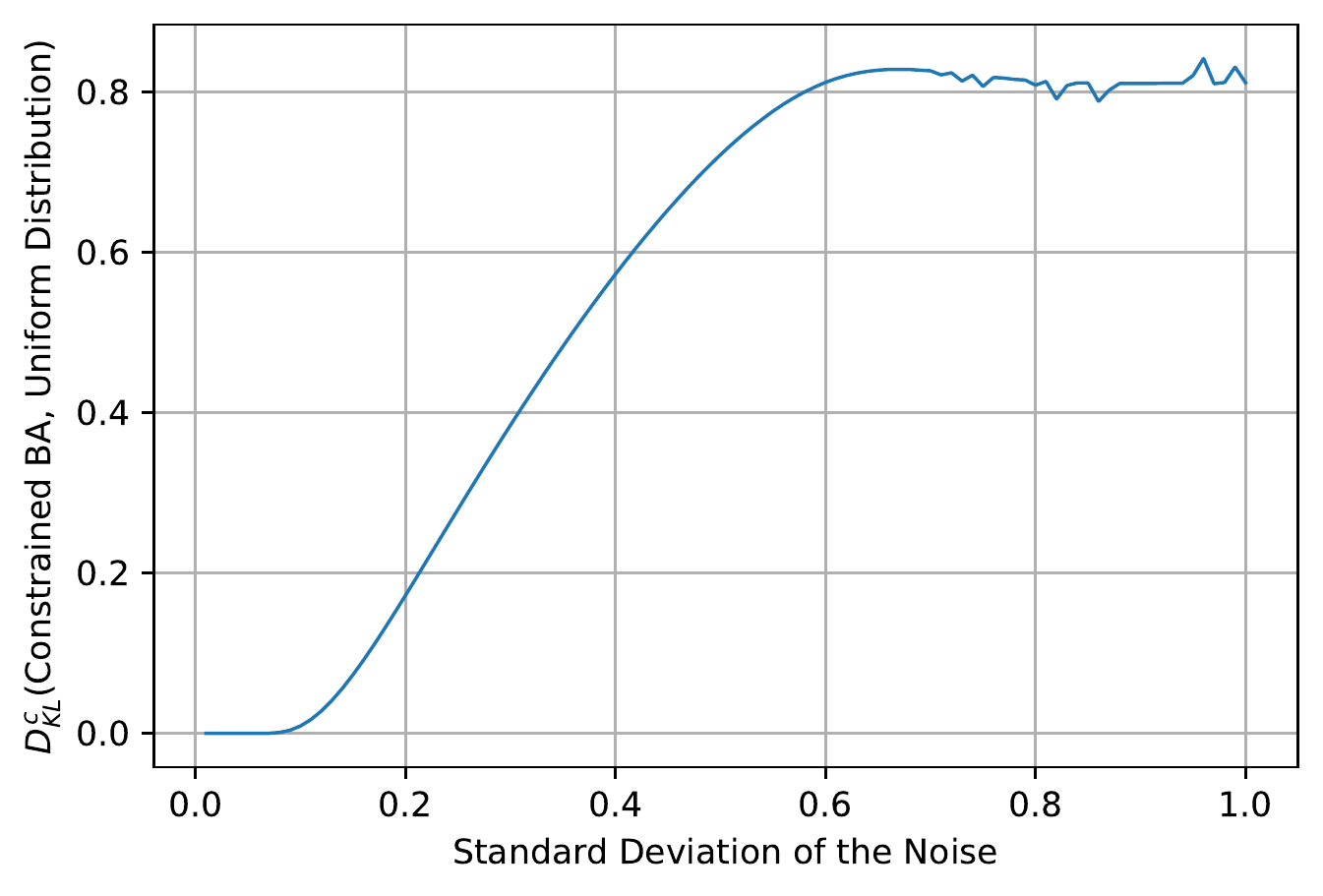}
        \caption[]%
        {{\small Commutative KL divergence between the uniform distribution and the constrained BA}}    
        \label{fig:CBA KL unif}
    \end{subfigure}
    \hfill
    \begin{subfigure}[b]{0.45\textwidth}  
        \centering 
        \includegraphics[width=\textwidth]{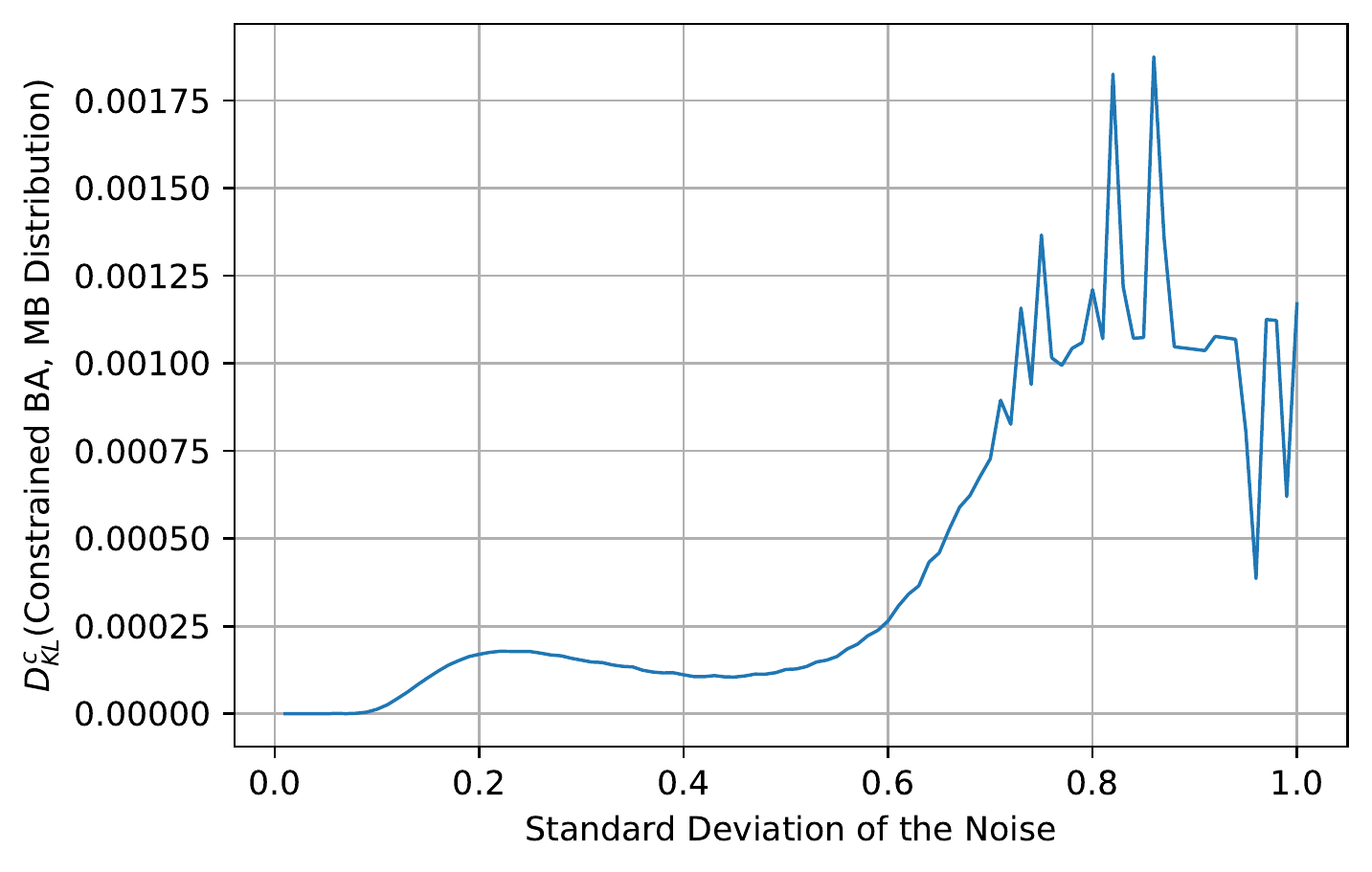}
        \caption[]%
        {{\small Commutative KL divergence between the constrained BA and the MB distribution}}    
        \label{fig:CBA KL MB}
    \end{subfigure}
    \caption{Divergence of the input probability distributions}
    \label{fig:CBA KL}
\end{figure*}

\begin{figure}[ht]
    \centering
    \includegraphics[width=9cm]{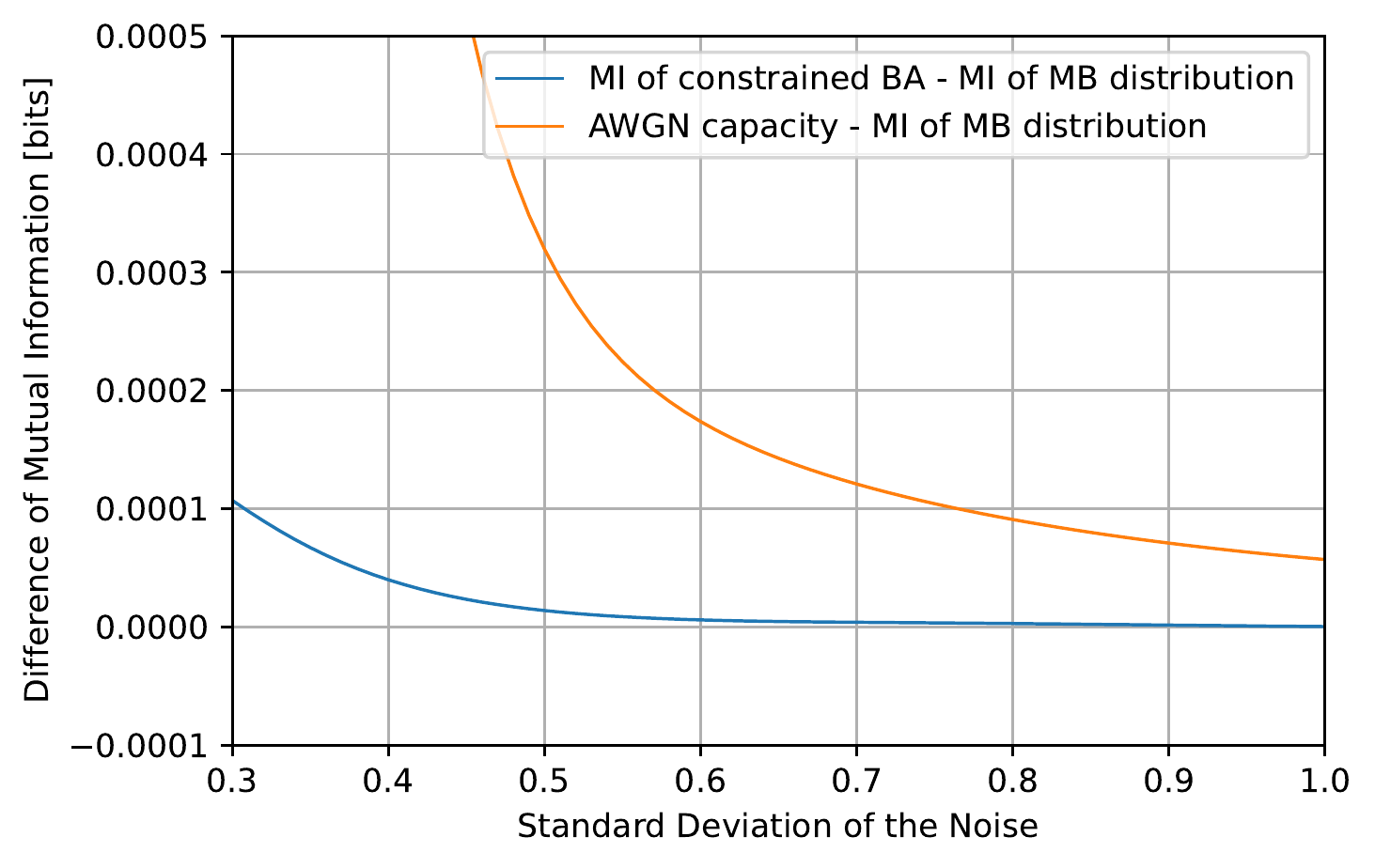}
    \caption{Mutual information given by the constrained Blahut--Arimoto and the AWGN capacity minus the mutual information given by the Maxwell--Boltzmann distribution}
    \label{fig:8-PAM CBA-MB vs AWGN-MB}
\end{figure}

In order to make pertinent comparisons, we will increase the precision of the constrained Blahut--Arimoto algorithm. The stopping criterion for the modified Blahut--Arimoto is now set to 1e-7, the depth of the search is set to twenty and the number of iterations per depth is set to fifty.\\\\
In Figure~\ref{fig:8-PAM CBA, MB, UNIF over SNR} is shown the mutual information over the SNR of a 8-PAM constellation. The figure compares the mutual information achieved by the constrained Blahut--Arimoto with the uniform distribution and the Maxwell--Boltzmann distribution.
\begin{figure}[ht]
    \centering
    \includegraphics[width=10cm]{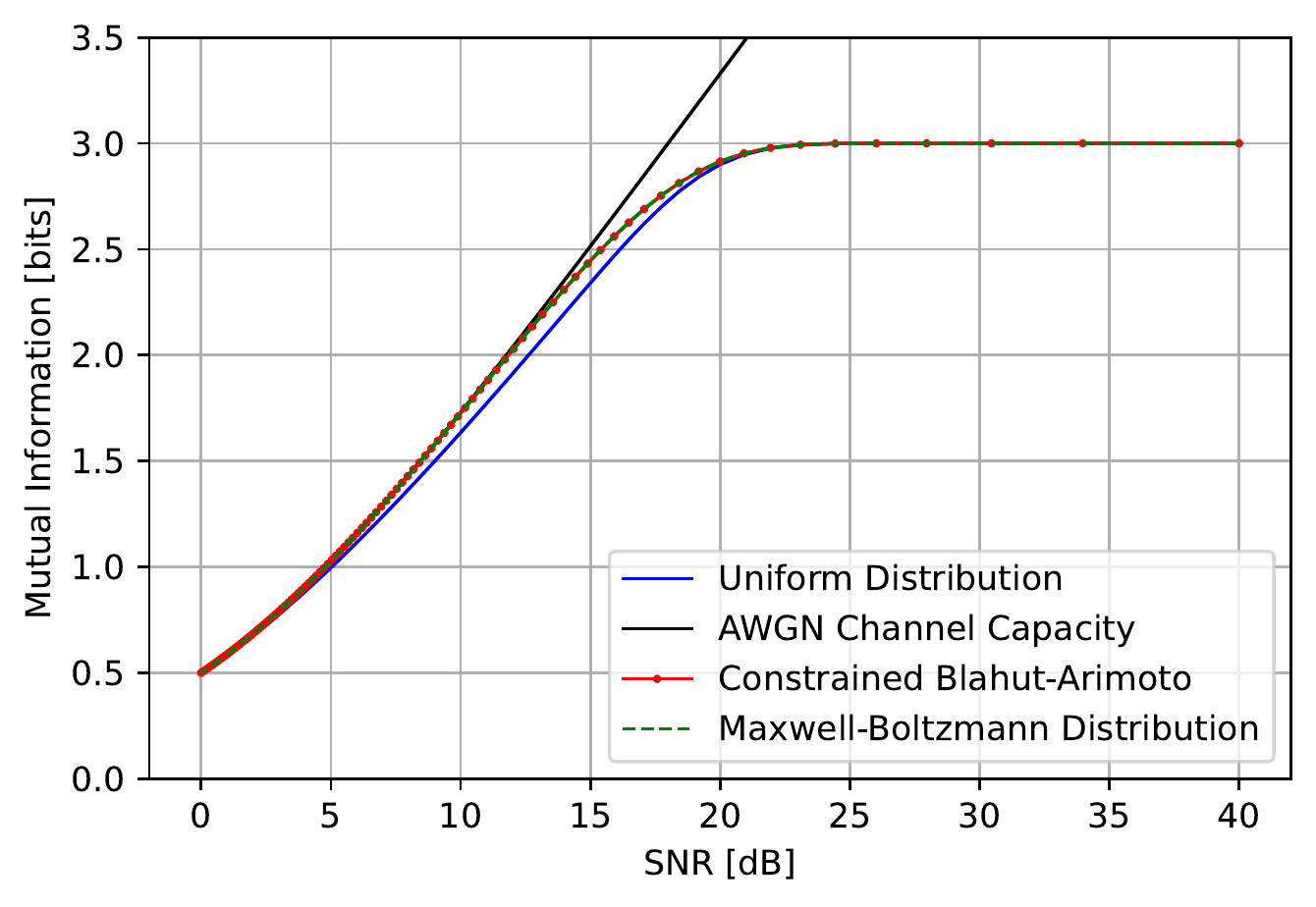}
    \caption{MI of a 8-PAM constellation over SNR given by different distributions}
    \label{fig:8-PAM CBA, MB, UNIF over SNR}
\end{figure}
As expected the constrained Blahut--Arimoto yields a bigger mutual information than the one given by the uniform distribution. We also notice that the constrained Blahut--Arimoto curve is very similar to the curve given by the Maxwell--Boltzmann distribution. Indeed, the energy of the difference between the two curves is approximately $4.6\cdot10^{-7}$.\\
We will now take a look, in Figure~\ref{fig:CBA diff}, at the difference in mutual information between the curves over the standard deviation of the noise.
\\
In Figure~\ref{fig:CBA diff unif} we see that, it makes a big difference if we use the constrained Blahut--Arimoto for higher standard deviations instead of the uniform distribution. This is not surprising as for a high $\sigma$, the optimal input distribution is the uniform one. So the constrained Blahut--Arimoto converges to the uniform distribution as the noise decreases. We will confirm this later by looking at the KL divergence between the uniform and the constrained Blahut--Arimoto distributions.\\
Concerning Figure~\ref{fig:CBA diff MB} we see that, the mutual information from the constrained Blahut--Arimoto is never lower than the one from Maxwell--Boltzmann distribution. We could expect this result since the constrained Blahut--Arimoto \textit{is by construction} the algorithm that gives the maximum mutual information for a given constellation. Note that to plot this graphics, we used a very thin resolution in terms of the thresholds and in terms of the number of iterations chosen. If we chose more imprecise parameters, we can expect that the difference of mutual information goes, for some standard deviations, under zero. Though, it would stay in a range of the horizontal axis that depends on the thinness of the parameters. Finally, we notice that there is a spike at $\sigma = 0.2$ which means that, especially for this standard deviation, it can be interesting to use the constrained Blahut--Arimoto instead of the Maxwell--Boltzmann distribution. Though we have to keep in mind that, in the absolute, the difference is very low. We are talking about a difference of at most 2e-4 bits per channel use.
\\\\
We will now have a look at the KL divergence to see if the distributions given by the different algorithms are similar or not. Figure~\ref{fig:CBA KL} shows the commutative KL divergence $D_{KL}^c$ over the standard deviation of the noise $\sigma$.
\\
In Figure~\ref{fig:CBA KL unif}, we see that for very low standard deviations, the probability distribution given by the constrained Blahut--Arimoto algorithm is very similar to the uniform distribution. Then, as the energy of the noise increases, the constrained Blahut--Arimoto diverges from the uniform distribution until reaching a plateau at $\sigma = 0.6$. This confirms that the uniform distribution is optimal for low noise energy.\\
Concerning the KL distance with the Maxwell--Boltzmann distribution, we notice interesting things. Overall, the KL divergence is very low compared to the divergence between the uniform distribution and the constrained BA. In Figure~\ref{fig:CBA diff MB}, we have noticed that there is a spike at $\sigma = 0.2$. Surprisingly, in \ref{fig:CBA KL MB}, the divergence at $\sigma = 0.2$ is not the highest spike at all. Indeed, from $\sigma = 0.6$ we can see a higher divergence. This means that, for high standard deviations, the two distributions have a different way of achieving the same mutual information. We can also conclude that, a relatively small change in the distribution given by Maxwell--Boltzmann leads to a relatively big improvement in mutual information around $\sigma = 0.2$.
\\\\
Finally, we know that the curve given by the constrained Blahut--Arimoto is between the one given by the Maxwell--Boltzmann distribution and the curve of the AWGN capacity. So we can ask ourselves which curve is closer to the constrained Blahut--Arimoto. To answer this question, we look at the plot in Figure~\ref{fig:8-PAM CBA-MB vs AWGN-MB} that shows the AWGN capacity minus the Maxwell--Boltzmann distribution. It also shows the curve of the constrained Blahut--Arimoto minus the Maxwell-Boltmann distribution.
\\
As we can see, the curve of the constrained Blahut--Arimoto is clearly closer to the horizontal axis than to the AWGN capacity for all standard deviations of the noise. This means that the mutual information obtained from the constrained Blahut--Arimoto is closer to the mutual information given by the Maxwell--Boltzmann distribution.

%% file: Chapters/7Conclusion.tex
\chapter{Conclusion}

The goal of the semester project was to learn about the concept of shaping. To this end, we first computed the curves of the mutual information under a uniform input distribution for the AWGN channel. We did it for one-dimensional and two-dimensional constellations. We then used the Maxwell--Boltzmann distribution to get a better mutual information. We were able to get a curve that gets close to the AWGN capacity. But, we also saw that the Maxwell--Boltzmann distribution is not the optimal distribution. Indeed, it maximises only the entropy of the input, with a power constraint, and not the mutual information. We used the Blahut--Arimoto algorithm to get the optimal distribution for a fixed constellation. The distribution obtained is optimal for a given noise energy but not for a given SNR.
Finally, we implemented a variation of the Blahut--Arimoto algorithm that adds a gain to the input. By doing so, we were able to get the optimal input distribution for a given SNR. We observed, as expected, that the curve of the mutual information obtained with this algorithm is between the curve given by the Maxwell--Boltzmann distribution and the curve of the AWGN capacity.\\
To summarise, the distribution obtained by Maxwell--Boltzmann is not optimal. However, we have seen that it achieves a mutual information very close to the one obtained from the Blahut--Arimoto algorithm if the temperature parameter is non-positive, for a given noise energy. Similarly, if the temperature parameter is non-negative and the SNR is fixed, we get a good approximation of the optimal distribution from Maxwell--Boltzmann.\\
As we were able to implement an algorithm that computes the capacity of a given constellation, we can consider that the goal of the project is achieved. The next steps, would be to learn about geometric shaping. We could also try to design an algorithm that uses both geometric and probabilistic shaping as a mathematical exercise.